\documentclass[12pt]{article}
\usepackage[bookmarks=true]{hyperref}
\usepackage{geometry}
\usepackage{graphicx}
\usepackage{amssymb}
\usepackage{amsmath}
\usepackage{amsfonts}
\usepackage{mathrsfs}
\usepackage{bbm}

\newcommand{\sSO}{SO}
\newcommand{\CN}{\mathcal{N}}
\newcommand{\di}{\mathrm{i}}   
\newcommand{\unit}{\mathbb{I}}   
\def\slasha#1{\setbox0=\hbox{$#1$}#1\hskip-\wd0\hbox to\wd0{\hss\sl/\/\hss}}

\newcommand{\de}{e}
\newcommand{\dpar}{\partial}
\newcommand{\sSU}{ SU}
\newcommand{\dd}{\mathrm{d}}
\newcommand{\tr}{{\rm tr}}
\newcommand{\CO}{\mathcal{O}}
\newcommand{\CCD}{\mathscr{D}}

\newcommand{\CF}{\mathcal{F}}
\newcommand{\CG}{\mathcal{G}}

\newcommand{\CU}{\mathcal{U}}
\newcommand{\CZ}{\mathcal{Z}}
\newcommand{\CX}{\mathcal{X}}
\newcommand{\CY}{\mathcal{Y}}
\newcommand{\CQ}{\mathcal{Q}}
\newcommand{\CB}{\mathcal{B}}
\newcommand{\der}[1]{\frac{\dpar}{\dpar #1}}   		
\newcommand{\CA}{\mathcal{A}}    			
\newcommand{\au}{u}
\newcommand{\asu}{su}
\newcommand{\CM}{\mathcal{M}}
\newcommand{\FR}{\mathbbm{R}}     			

\newcommand{\nablas}{\slasha{\nabla}}
\newcommand{\CV}{\mathcal{V}}
\newcommand{\CCV}{\mathscr{V}}
\newcommand{\bth}{{\bar{\theta}}}
\newcommand{\thetas}{{\slasha{\theta}}}
\newcommand{\eand}{\text{\ and\ }}  
\newcommand{\sU}{ U} 
\newcommand{\eps}{\epsilon}

\usepackage{todonotes}

\title{\vspace*{-0.5cm}\bf\LARGE Super Yang-Mills Theory with Impurity Walls and Instanton Moduli Spaces
}
\author{\Large 
Sergey A. Cherkis$^\flat$, Clare O'Hara$^\natural$, and Christian S\"amann$^\sharp$\\
\\
$^{\flat,\natural}$\it \normalsize School of Mathematics,  Trinity College, Dublin, Ireland\\
$^\flat$\it\normalsize Hamilton Mathematics Institute,  Trinity College, Dublin, Ireland,\\
\it \normalsize Department of Mathematics, Stanford University, CA 94305, {\rm and}\\
 \it \normalsize Department of Physics, University of California, Berkeley, CA 94720\\
$^\sharp$\it \normalsize Maxwell Institute for Mathematical Sciences, Edinburgh, UK\\
\normalsize  and \it Department of Mathematics, Heriot-Watt University\\
\it \normalsize Colin Maclaurin Building, Riccarton, Edinburgh EH14 4AS, UK\\
\small \tt cherkis@maths.tcd.ie, clohara@maths.tcd.ie, c.saemann@hw.ac.uk
}
                                         
\begin{document}
\begin{titlepage}

\renewcommand{\thepage}{ }
\date{}

\maketitle
\vspace*{-0.7cm}
\abstract{We explore maximally supersymmetric Yang-Mills theory with walls of impurities respecting half of the supersymmetries.  The walls carry fundamental or bifundamental matter multiplets.  We employ three-dimensional $\CN=2$ superspace language to identify the Higgs branch of this theory.  We find that the vacuum conditions determining the Higgs branch are exactly the bow equations yielding Yang-Mills instantons on a multi-Taub-NUT space.  

Under electric-magnetic duality, the super Yang-Mills theory describing the bulk is mapped to itself, while the fundamental- and bifundamental-carrying impurity walls are interchanged.  We perform a one-loop computation on the Coulomb branch of the dual theory to find the asymptotic metric on the original Higgs branch.} 

\vspace{-7.5in}

\parbox{\linewidth}
{\small\hfill \shortstack{
\hfill TCDMATH 11-01, HMI 11-01\\
\hfill EMPG-10-28, HWM-10-37
}}

\end{titlepage}

\tableofcontents

\section{Introduction}
A string theory realization of a quantum gauge theory can be very useful in analyzing the latter \cite{Elitzur:1997fh, Klemm:1996bj, Witten:1997sc, Maldacena:1997re}.  Such a realization can relate an intrinsically quantum problem in the gauge theory to a different amenable classical problem.  For example, three-dimensional quantum gauge theories were related to the classical  dynamics of monopoles in \cite{Seiberg:1996nz,Chalmers:1996xh} and singular monopoles in \cite{Cherkis:1997aa}.  In this work, however, we employ such a string theory realization in a reverse fashion:  We study a supersymmetric quantum gauge theory to make statements about the classical moduli space of Yang-Mills instantons on multi-Taub-NUT space with $k$ centers ($TN_k$).

The gauge theory we are interested in is four-dimensional and possesses\footnote{To avoid confusion, we give the space-time dimension explicitly as a subscript. For example, a four-dimensional theory with $\CN_{\rm 4d}=2$ has the same number of conserved real supercharges as a three-dimensional theory with $\CN_{\rm 3d}=4$.} $\CN_{\rm 4d}=2$ supersymmetry. In the four-dimensional bulk, the theory is given by maximally supersymmetric, i.e.\ $\CN_{\rm 4d}=4,$ Yang-Mills theory.  Half of the bulk supersymmetry is broken by the presence of codimension one defects.  The defect walls are all parallel, each carrying either a fundamental or a bifundamental $\CN_{\rm 3d}=4$ hypermultiplet.  We choose the gauge group in  each space bounded by the bifundamental-carrying impurity walls to be unitary.  The rank of the gauge group can a priori differ on the two sides of such a wall.

Without the bifundamental defects, such a theory was studied in \cite{Kapustin:1998pb, DeWolfe:2001pq,Erdmenger:2002ex}.  As in \cite{Erdmenger:2002ex}, we split four-dimensional space into a direct product of a three-dimensional space parallel to the defects and a one-dimensional space perpendicular to these. The latter is parameterized by the coordinate $s=x^6$. To work in a superspace framework, we  embed $\CN_{\rm 3d}=2$ superspace in a compatible way into $\CN_{\rm 4d}=2$ superspace. Our gauge theory is therefore formulated in terms of $\CN_{\rm 3d}=2$ superfields on $\FR^{1,2}$ that depend on $s$ as a parameter. Correspondingly, the vacuum D- and F-flatness conditions, which are usually algebraic, now take the form of differential equations in this variable $s$.  

We find that on the Higgs branch of the gauge theory these equations are exactly the bow equations of \cite{Cherkis:2009jm} which describe $U(n)$ instantons on a multi-Taub-NUT space, $TN_k.$ The number of the Taub-NUT centers $k$ equals the number of the bifundamental impurity walls of the gauge theory, while the rank $n$ of the instanton gauge group equals the number of the fundamental impurity walls.  The instanton charges are determined by the ranks of the gauge group of the gauge theory.  This identifies the Higgs branch of the theory which has $n$ fundamental and $k$ bifundamental impurity walls with the moduli space of $U(n)$ instantons on $TN_k.$    Obtaining these differential equations from the D- and F-flatness conditions and relating them to the instanton bow data is one of the results of this work.  The equations we obtain, however, are slightly more general and suggest an interpretation of the instanton problem as a part of a larger framework provided by the quantum gauge theory.  

Once we identify the Higgs branch of our impurity theory with the moduli space of instantons on $TN_k,$ we acquire an entirely new way of computing the metric on the latter: We can apply electric-magnetic duality \cite{Montonen:1977sn, Witten:1978mh} (which in three-dimensional language corresponds to gauge theory mirror symmetry \cite{Intriligator:1996ex, Kapustin:1999ha}), so that instead of considering the Higgs branch of our original theory, we study the same space as the Coulomb branch of the dual theory. In particular, the asymptotic form of the moduli space metric is determined by perturbative corrections to the propagator of the gauge theory.  Moreover, on the Coulomb branch  the theory is effectively three-dimensional in the extreme infrared.  We compute the one-loop correction in this three-dimensional theory, generalizing the results of \cite{Dorey:1997ij,Dorey:1998kq}. This leads exactly to the Lee-Weiberg-Yi-type \cite{Lee:1996kz} asymptotic metric, which was found to be the asymptotic metric of the instanton moduli space in \cite{Cherkis:2010bn}.

The underlying relation of the moduli space of vacua of the impurity gauge theory to the moduli space of instantons is not coincidental: A string theory realization of our impurity gauge theory is given by a Chalmers-Hanany-Witten (CHW) brane configuration of D5-, NS5-, and D3-branes in type IIB string theory \cite{Chalmers:1996xh, Hanany:1996ie, Cherkis:2008ip}. Applying T-duality along the D3 relative-transverse  direction parameterized by $s$, one maps this configuration to a type IIA string theory configuration with $n$ D6-branes wrapped on $TN_k$ and some D2-branes transverse to $TN_k$. The latter can be  argued \cite{Douglas:1995bn,Douglas:1996sw} to be effectively described by instantons. 

Solutions of type IIB supergravity corresponding to fully back-reacted geometries of $AdS_5\times S^5$ with probe D5- and NS5-branes have been found in \cite{D'Hoker:2007xy,D'Hoker:2007xz}. These solutions are supersymmetric versions of Janus solutions \cite{Bak:2003jk,Clark:2004sb} and possess $SO(2,3)\times SO(3)\times SO(3)$ symmetry. Both their geometric interpretation and their symmetry group suggest a close connection to CHW configurations. In fact it has been argued \cite{D'Hoker:2007xz} that the gauge theory dual is one of the so called interface theories \cite{Clark:2004sb}, which were classified in \cite{D'Hoker:2006uv}: a four-dimensional supersymmetric gauge theory with defect walls. Contrary to our impurity theories, these interface theories do not come with additional degrees of freedom on the impurity walls.  It would be very interesting to study the relation between these AdS solutions and the impurity theory considered here in more detail.

Our paper is structured as follows: Having derived the matter content from the analysis of the CHW configuration in Section~\ref{Sec:CHW}, we give the gauge theory Lagrangian in Section~\ref{Sec:gaugeTheoryAction}. The vacuum conditions are discussed in Section~\ref{Vacuum}, where we also compare them to the instanton data. Section~\ref{Sec:OneLoop} contains the calculation of the one-loop quantum corrections to the metric on the moduli space, and we conclude in Section~\ref{sec:Conclusion}. Our  conventions are summarized in the appendix.

\section{The Chalmers-Hanany-Witten brane configuration and instantons}\label{Sec:CHW}

In order to make various key ingredients in our discussion transparent, let us begin with the description of a Chalmers-Hanany-Witten D-brane configuration \cite{Chalmers:1996xh,Hanany:1996ie}. This configuration is the type IIB string theory background summarized in Table~\ref{Table:Eng}.  Its background geometry is ten-dimensional Minkowski space with one spatial dimension compactified on a circle: $\mathbb{R}^{1,2}\times\mathbb{R}^3_Z\times S^1\times\mathbb{R}^3_Y$. As coordinates on the various components of this product space, we use
$(x_0,x_1,x_2)\in\mathbb{R}^{1,2},\ \vec{z}\in\mathbb{R}^3_Z,\ s\in S^1$ and $\vec{y}\in\mathbb{R}^3_Y.$
The space contains $n$ parallel D5-branes with world-volumes $\mathbb{R}^{1,2}\times\mathbb{R}^3_Y$ located at\footnote{Our notation is chosen  to match that of \cite{Cherkis:2010bn} later in the discussion.} $\lambda_j\in S^1$, $j=1,\ldots,n$ and at the origin of $\mathbb{R}^3_Z,$ i.e.\ at  $\vec{z}=0$.  We also have $k$ distinct parallel NS5-branes with world-volumes  $\mathbb{R}^{1,2}\times\mathbb{R}^3_Z$ positioned at $p_\sigma\in S^1$ and $\vec{\nu}_\sigma\in\mathbb{R}^3_Y$, $\sigma=1,\ldots,k$. The last important ingredient is  a collection of D3-branes that are either suspended between D5-branes, having world-volumes $\mathbb{R}^{1,2}\times[\lambda_j, \lambda_{j+1}]$, or wrap the circle factor entirely and have  world-volume $\mathbb{R}^{1,2}\times S^1.$  

Such configurations were thoroughly analyzed using their effective description in terms of three-dimensional gauge theories and mirror symmetry \cite{Intriligator:1996ex, Aharony:1997bx}.  They also proved to be very useful in the exploration of singular monopoles \cite{Cherkis:1997aa,Cherkis:1998hi}.  Here we focus on the effective gauge theory in the {\em four-dimensional  world-volume} of the D3-brane describing its - still four-dimensional - low energy dynamics. 

Away from any five-branes, the effective infrared description of the CHW configuration is given by maximally supersymmetric Yang-Mills theory. The presence of the five-branes manifests itself in the form of two types of defects in this theory.  It is important to emphasize the different geometric nature of the two kinds of defect walls that we consider.  A fundamental wall, i.e.\ a defect wall carrying a fundamental matter multiplet, is contained within the four-dimensional space-time.  A gauge transformation in the bulk acts on the fundamental multiplet by its value at the wall.  A bifundamental defect  positioned at $s=p_\sigma,$ on the other hand, separates the space-time into a half-space-time $s\geq p_\sigma$ to its right  and another half-space-time $s\leq p_\sigma$ to its left.   

In order to make this transparent, we introduce two distinct points $p_{\sigma}^L$ and $p_{\sigma-1}^R$ which are located at the boundary of the $s$-semi-axis in the space on the right and on the left of the wall, respectively. With these conventions, any field $\phi$ continuous on one side of the wall satisfies\footnote{We use $\lim_{p\nearrow p_{\sigma}}$ and $\lim_{p\searrow p_{\sigma}}$ to denote the left and right one-sided  limits, respectively.} $\lim_{p\nearrow p_{\sigma}}\phi(p)=\phi(p_{\sigma-1}^R)$ and  $\lim_{p\searrow p_{\sigma}}\phi(p)=\phi(p_{\sigma}^L).$ As a result, $k$ bifundamental walls separate the space-time into $k$ slices so that the $\sigma^\text{th}$ slice corresponds to the interval $[p_\sigma^L, p_\sigma^R].$ The gauge transformations act independently at $p_{\sigma-1}^R$ and at $p_\sigma^L.$  Because of the presence of $k$ walls with bifundamental multiplets, the bulk gauge theory is defined on $k$ independent slabs $\mathbb{R}^{1,2}\times[p_\sigma^L, p_\sigma^R]$. In order to deal with the boundary terms and integration by parts\footnote{See discussion on page \pageref{Page:GenDer}.} for the fields in a given slab $\mathbb{R}^{1,2}\times[p_\sigma^L, p_\sigma^R]$, we understand the fields to be extended by zero on $\mathbb{R}^{1,2}\times[p_\sigma^L-\epsilon, p_\sigma^R+\epsilon]$.

Altogether, the gauge theory action $S=S_{\rm bulk}+S_f+S_b$ and consists of
\begin{itemize}
\renewcommand{\labelitemi}{$\cdot$}
\item the bulk contribution $S_{\rm bulk},$ given by the action of maximally supersymmetric Yang-Mills theory written in terms of $\CN_{\rm 3d}=2$ superfields, 
\item the fundamental multiplet contribution $S_f$ which, since these multiplets are localized at $s=\lambda_j,$ has a three-dimensional Lagrangian density containing couplings to the bulk fields values at these values of $s$,
and
\item the bifundamental contribution $S_b$ with the $\sigma^\text{th}$ term of its three-dimensional Lagrangian density containing couplings of the $\sigma^\text{th}$ bifundamental multiplet to the bulk fields at $p_{\sigma-1}^L$ and at $p_\sigma^R$. 
\end{itemize}
As a result, any variational equation obtained by varying a bulk field contains contributions from the $j^\text{th}$ fundamental multiplets with a factor $\delta(s-\lambda_j)$. Additionally, contributions from the $\sigma^\text{th}$ bifundamental multiplet appear with a factor of either $\delta(s-p_{\sigma-1}^L)$ or $\delta(s-p_\sigma^R).$ We present the action in Section~\ref{Sec:gaugeTheoryAction} and we discuss the variational equations in detail in Section~\ref{Vacuum}. 

\begin{table}[htdp]
\begin{center}
\begin{tabular}{|c|c|c|c|c|c|c|c|c|c|c|}
\hline
  & 0 & 1 & 2 & 3 & 4 & 5 & 6  & 7 & 8 & 9\\
\hline
Coordinates & $x^0$ & $x^1$ & $x^2$ & \multicolumn{3}{|c|}{$\vec{z}$} & $s$ & \multicolumn{3}{|c|}{$\vec{y}$} \\
\hline
Symmetries & \multicolumn{3}{|c|}{$\sSO(1,2)$} & \multicolumn{3}{|c|}{$\sSO(3)_Z$} & $ $ & \multicolumn{3}{|c|}{$\sSO(3)_Y$} \\
\hline
 NS5 & $\times$ & $\times$ & $\times$ & $\times$ & $\times$ & $\times$ & $p_\sigma$ & \multicolumn{3}{|c|}{$\vec{\nu}_\sigma$} \\
\hline
 D5 & $\times$ & $\times$ & $\times$ & \multicolumn{3}{|c|}{$\vec{0}$} & $\lambda_j$ & $\times$ & $\times$ & $\times$ \\
\hline
 D3 & $\times$ & $\times$ & $\times$ & \multicolumn{3}{|c|}{$\vec{0}$} & $\times$ & \multicolumn{3}{|c|}{$\vec{y}\,^{D3}$} \\
$\CN=1$  fields/$\Psi$ & \multicolumn{3}{|c|}{$\CV$} & \multicolumn{2}{|c|}{$\CZ$} & $\CV$ & & & \multicolumn{2}{|c|}{} \\
$\Psi$ components & $v_0$ & $v_1$ & $v_2$ & \multicolumn{2}{|c|}{$Z$} & $Z_3$ & &  & \multicolumn{2}{|c|}{} \\
$\CN=1$ fields/$\Upsilon$ & & & & \multicolumn{2}{|c|}{} & & \multicolumn{2}{|c|}{$\CX$} & \multicolumn{2}{|c|}{$\CY$} \\
$\Upsilon$ components& & & & \multicolumn{2}{|c|}{} & & $v_6$ & $Y_1$ & \multicolumn{2}{|c|}{$Y$} \\
\hline
\end{tabular}
\end{center}
\caption{The CHW brane configuration, its symmetries, and the components of the supermultiplets in its effective gauge theory description. Here, $Z=Z_1+\di Z_2$ and $Y=Y_2+\di Y_3$. The superfields $\Psi$ and $\Upsilon$ denote $\CN_{\rm 4d}=2$ vector- and hypermultiplets, cf.\ the appendix.}
\label{Table:Eng}
\end{table}%
A CHW brane configuration preserves 8 of the 16 real supercharges. That is, we expect the gauge theory to exhibit $\CN_{\rm 3d}=4$ supersymmetry with R-symmetry algebra
$so(4)\simeq su(2)_Z\times su(2)_Y\simeq so(3)_Z\times so(3)_Y$. The Higgs field triplets $(Z_1, Z_2, Z_3)$ and $(Y_1,Y_2, Y_3)$ form vector representations of the factors $so(3)_Z$ and $so(3)_Y$. These algebras correspond to rotations in the spaces $\FR^3_Z$ and $\FR^3_Y$ respectively.

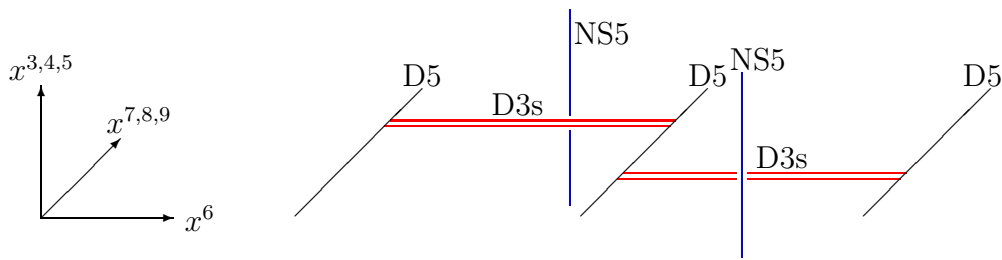
\begin{figure}[h]
\center
\begin{picture}(290,95)
\put(-30.0,15.0){\vector(1,0){50}}
\put(-30.0,15.0){\vector(0,1){50}}
\put(-30.0,15.0){\vector(1,1){30}}
\put(30.0,15.0){\makebox(0,0)[c]{$x^6$}}
\put(-30.0,72.0){\makebox(0,0)[c]{$x^{3,4,5}$}}
\put(7.0,52.0){\makebox(0,0)[c]{$x^{7,8,9}$}}
\put(170.0,20.0){\color{blue}\line(0,1){28}}
\put(170.0,54.0){\color{blue}\line(0,1){40}}
\put(102.0,52.0){\color{red}\line(1,0){108}}
\put(100.0,50.0){\color{red}\line(1,0){108}}
\put(188.0,30.0){\color{red}\line(1,0){45}}
\put(190.0,32.0){\color{red}\line(1,0){43}}
\put(237,30.0){\color{red}\line(1,0){58}}
\put(237,32.0){\color{red}\line(1,0){60}}
\put(174.0,16.0){\line(1,1){48}}
\put(281.0,16.0){\line(1,1){48}}
\put(66.0,16.0){\line(1,1){48}}
\put(235.0,0.0){\color{blue}\line(0,1){70}}
\put(182.0,85.0){\makebox(0,0)[c]{NS5}}
\put(241.0,75.0){\makebox(0,0)[c]{NS5}}
\put(222.0,69.0){\makebox(0,0)[c]{D5}}
\put(114.0,69.0){\makebox(0,0)[c]{D5}}
\put(326.0,69.0){\makebox(0,0)[c]{D5}}
\put(150.0,58.0){\makebox(0,0)[c]{D3s}}
\put(250.0,38.0){\makebox(0,0)[c]{D3s}}
\end{picture}
\caption{A picture of a Chalmers-Hanany-Witten configuration in the Y- or Higgs branch.}\label{fig:pictureCHW}
\end{figure}

Since the D3-branes can end on D5-branes if their positions in $\FR_Z^3$ agree, we can have a branch of the space of vacua of the theory where all of the D3-branes are positioned at $\vec{z}=0$ and any separation between the D3-branes is along the $\mathbb{R}^3_Y$ space factor.  We call this the {\em Y-branch} of the theory. Potentially, if any of the NS5-brane positions $\vec{\nu}_\sigma$ coincide, there is another branch, the {\em Z-branch}, with D3-branes ending on NS5-branes positioned at the associated values of $\vec{y}=\vec{\nu}_\sigma$ and arbitrary values of $\vec{z}.$  There is also a {\em mixed branch} corresponding to at least some of the D3-branes having world-volumes $\mathbb{R}^{1,2}\times S^1$ positioned at nonzero $\vec{z}$ and $\vec{y}$ or some D3-branes separated along the Z- while others are separated along the Y-directions.  
Here, we assume that all of the NS5-brane positions $\vec{\nu}_\sigma$ are distinct and therefore the Z-branch does not arise. An example brane configuration on the Y-branch is depicted in Figure \ref{fig:pictureCHW}.  
  
We deliberately name these branches according to the directions in which the D3-branes are separated from each other and not according to the types of the gauge theory supermultiplets that parameterize them.  This avoids potential confusion that can arise once the gauge theory mirror symmetry interchanging the roles of the supermultiplets enters the discussion.  Once we consider the gauge theory description of the low energy D3-brane dynamics in this CHW configuration, the Y-branch, which is the main object of our study, corresponds to the Higgs branch of the gauge theory.  The Z-branch, if it existed, would correspond to the Coulomb branch off that  gauge theory.  
In Section \ref{Sec:OneLoop}, however, where we perform perturbative computations at one loop, we shall find it convenient to work in the mirror or S-dual picture.  We still study the Y-branch, but once the mirror symmetry is applied, the Y-branch is identified with the Coulomb branch of the mirror theory.  It is the one-loop computation involving the vector 
multiplet that gives us the asymptotic metric on the Y-branch.

This brane configuration is directly related to Yang-Mills instantons as follows. Whenever the $x^6$ direction is compact, we can T-dualize along it. This yields a dual type IIA string theory brane configuration which contains the $k$-centered Taub-NUT space in the $6,7,8,9$-directions, a number of D6-branes wrapping it with their world-volumes in the $0,1,2,6,7,8,9$ directions, and D2-branes (some of them fractional) with world-volumes extending in the $0,1,2$ directions.  Because of a modification of the argument of \cite{Douglas:1995bn,Douglas:1996sw}, this configuration is described by instantons on the space $TN_k$.  If the $x^6$ direction is noncompact or if some $x^6$ intervals have no D3-branes over them, the corresponding instanton on $TN_k$ is independent of the isometric direction (i.e.\ the direction T-dual to $x^6$) and can be interpreted as a singular monopole. The relation between CHW configurations and instantons on a multi-Taub-NUT space was used in \cite{Cherkis:2009jm,Cherkis:2008ip} to describe such instanton configurations and their moduli spaces explicitly. Singular monopoles were constructed using this interpretation in \cite{Blair:2010kz,Blair:2010vh}.

\section{Gauge theory action}\label{Sec:gaugeTheoryAction}

\subsection{Field content}

The bosonic degrees of freedom in the $(x_0,x_1,x_2,x_6)$ bulk of the D3-brane are the gauge field $(v_0,v_1,v_2,v_6),$ three antihermitian Higgs fields $Z_1,Z_2,Z_3$ (corresponding to the relative-transverse directions along the $\mathbb{R}^3_Z$ component) and three hermitian Higgs fields $Y_1,Y_2,$ and $Y_3$ (corresponding to the relative-transverse directions along the $\mathbb{R}^3_Y$ component). They form the bosonic part of an $\CN_{\rm 4d}=4$ vector supermultiplet.

This $\CN_{\rm 4d}=4$ vector supermultiplet decomposes into an $\CN_{\rm 4d}=2$ vector multiplet $\Psi$ and an $\CN_{\rm 4d}=2$ hypermultiplet $\Upsilon$.  In $\CN_{\rm 4d}=1$ language, the $\CN_{\rm 4d}=4$ vector supermultiplet  contains an $\CN_{\rm 4d}=1$ vector supermultiplet and three $\CN_{\rm 4d}=1$ chiral supermultiplets. For our discussion, it will be useful to switch to $\CN_{\rm 3d}=2$ superspace.   We use the same embedding of $\CN_{\rm 3d}=2$ superspace into $\CN_{\rm 4d}=2$ superspace as in \cite{Erdmenger:2002ex}; the details are provided in the appendix for completeness.

We arrange the bosonic fields listed above as follows into a vector superfield $\CV$ (which gives rise to the complex linear superfield $\Sigma$) and three chiral superfields $\CX, \CZ,$ and $ \CY$;
\begin{equation}
 \begin{aligned}
    \CV&:(v_0,v_1,v_2,Z_3,\lambda,D)~,\\
    \CX&:(X,\psi,G)~,&\text{with}\ &X=v_6+\di Y_1~,\\
    \CU^1:=\CZ&:(Z,\chi^1,F^1)~,&\text{with}\ &Z=Z_1+\di Z_2~,\\
    \CU^2:=\CY&:(Y,\chi^2,F^2)~,&\text{with}\ &Y=Y_2+\di Y_3~.
 \end{aligned}
\end{equation}
Explicitly, the superfield expansions are
\begin{equation}
\begin{aligned}
 \CV&=\di\theta_\alpha\bth^\alpha Z_3-\theta\sigma^\mu_{\rm 3d}\bth v_\mu+\di\theta^2\bth\bar{\lambda}-\di\bth^2\theta\lambda+\tfrac{1}{2}\theta^2\bth^2D~,\\
 \CX&=v_6(y)+\di Y_1(y)+\sqrt{2}\theta\psi(y)+\theta^2 G(y)~,\\
 \CU^1&=\CZ=Z(y)+\sqrt{2}\theta\chi^i(y)+\theta^2 F^1(y)~,\\
 \CU^2&=\CY=Y(y)+\sqrt{2}\theta\chi^i(y)+\theta^2 F^2(y)~.
\end{aligned}
\end{equation}
Here, $\sigma^\mu_{\rm 3d}$ is the set of sigma matrices reduced to three dimensions. We use the convention $(\sigma^\mu_{\rm 3d})=(-\unit,\sigma^1,\sigma^3)$. The complex scalar fields $D,G,F^1$ and $F^2$ are auxiliary fields and $\lambda,\psi$ and $\chi$ are Weyl spinors in four dimensions. The four-dimensional gauge field $(v_0,v_1,v_2,v_6)$ is spread over the two $\CN_{\rm 3d}=2$ superfields $\CV$ and $\CX$. Correspondingly, the real scalar field $Z_3$, which might be expected to be a part of the $\CN_{\rm 3d}=2$ chiral superfield $\CX$, is a part of the vector superfield, as explained in \cite{Erdmenger:2002ex}.

We use the superfield conventions of \cite{Wess:1992cp}. In particular, we have 
\begin{equation}
 \theta\lambda:=\theta^\alpha\lambda_\alpha,~~~\bth\bar{\lambda}:=\bth_\alpha\bar{\lambda}^\alpha~~~\eand~~~ \theta\sigma^\mu\bar{\lambda}=\theta^\alpha\sigma^\mu_{\alpha\beta}\bar{\lambda}^\beta.
\end{equation}
For products of spinors of the form $\bth\lambda$, we always make the index structure explicit. A useful relation is
\begin{equation}
 (\bth_\alpha \theta^\alpha)^2=\tfrac{1}{2}\theta^2\bth^2~.
\end{equation}
We denote the adjoint of a field $\lambda$ by $\bar{\lambda}$ instead of $\lambda^\dagger$ to simplify notation.

Note that $v_6$ is a gauge field along the direction $x^6=s,$ and each $\CN_{\rm 3d}=2$ superfield has to be understood as depending on the parameter $s$. Gauge transformations  act on the superfields according to
\begin{equation}
 \begin{aligned}
    \CU^i&\rightarrow e^{-2\di\Lambda}\CU^i e^{2\di\Lambda}~,\\
    \CX&\rightarrow e^{-2\di\Lambda}\CX e^{2\di\Lambda}+ e^{-2\di\Lambda}\partial_{s} e^{2\di\Lambda}~,
 \end{aligned}
\end{equation}
where $\Lambda$ is an $\CN_{\rm 3d}=2$ chiral superfield depending on the parameter $s$. In three dimensions, the vector superfield $ \CV$ gives rise to the complex linear superfield 
\begin{equation}
 \Sigma:=\epsilon^{\alpha\beta}\bar D_\alpha(e^{2\di \CV}D_\beta e^{-2\di \CV})~.
\end{equation}
Its field expansion is
\begin{equation}
\begin{aligned}
 \Sigma(x,\theta,\bth)=&4Z_3-4\theta^\alpha\bar{\lambda}_\alpha-4\bth_\alpha\lambda^\alpha-4\di\bth_\alpha\theta^\alpha D-2\theta\sigma_{\rm 3d}^\mu\bth\eps_{\mu\nu\kappa}F^{\nu\kappa}\\
&~~-2\di\bth^2\theta\sigma^\mu_{\rm 3d}\nabla_\mu\lambda+2\di\theta^2\bth\bar{\sigma}^\mu_{\rm 3d}\nabla_\mu\bar{\lambda}-\theta^2\bth^2\nabla_\mu\nabla^\mu Z_3~.
\end{aligned}
\end{equation}
Since gauge transformations act on the vector superfield as
\begin{equation}
  e^{2\di \CV}\rightarrow e^{-2\di\Lambda} e^{2\di \CV}\de^{2\di\bar\Lambda}~,
\end{equation}
the complex linear superfield transforms according to
\begin{equation}
  \Sigma\rightarrow e^{-2\di\Lambda}\Sigma\de^{2\di\Lambda}~.
\end{equation}

\subsection{Action in superspace}

The following bulk action is manifestly invariant under both $\CN_{\rm 3d}=2$ supersymmetry and gauge symmetry:
\begin{equation}\label{actionbulk}
\begin{aligned}
 S_{\rm bulk}=\int \dd s\, \dd^3 x\,{\rm tr}\Big[&\int \dd^4\theta \left(-\tfrac{1}{16}\Sigma^2-\tfrac{1}{4}(\de^{2\di \CV}(\partial_s-\bar{\CX}) e^{-2\di \CV}-\CX)^2+\tfrac{1}{2}\de^{2\di \CV}\bar{\CU_i} e^{-2\di \CV}\CU^i\right)\\ &+\tfrac{\di}{2}\int  \dd^2\theta~\epsilon_{ij}\CU^i[\partial_s+\CX,\CU^j]-\tfrac{\di}{2}\int  \dd^2\bar{\theta}~\epsilon_{ij}\bar{\CU}^i[\partial_s-\bar{\CX},\bar{\CU}^j]\Big]~.
\end{aligned}
\end{equation}
Performing the Gra{\ss}mann integrals, one obtains the component action of $\CN_{\rm 4d}=4$ super Yang-Mills theory. 

As we used four-dimensional superfields, there is only a $U(1)$ subgroup  of the $SO(3)_Z\times SO(3)_Y$ R-symmetry group manifest. In component fields however, the full symmetry group is recovered after integrating out all auxiliary fields. There is an additional $\sSU(2)$ flavor symmetry acting on the doublet $(\CZ,\CY)$. This symmetry partially mixes fields belonging to the $\CN_{\rm 3d}=4$ vector and hypermultiplets.

Varying the action \eqref{actionbulk} with respect to the auxiliary fields yields
\begin{align}\label{Eq:Dbl}
\frac{\delta S_{\rm bulk}}{\delta D}=& D-\partial_s Y_1-[v_6,Y_1]-\tfrac{\di}{2}([Z,\bar{Z}]+[Y,\bar{Y}])~,\\
\label{Eq:Gbl}
\frac{\delta S_{\rm bulk}}{\delta G}=& -\tfrac{1}{2}\bar{G}-\di[Y,Z]~,\\
\label{Eq:F1bl}
\frac{\delta S_{\rm bulk}}{\delta F^1}=& \tfrac{1}{2}\bar{F}_1-\di\partial_sY-\di[v_6+\di Y_1,Y]~,\\
\label{Eq:F2bl}
\frac{\delta S_{\rm bulk}}{\delta F^2}=& \tfrac{1}{2}\bar{F}_2+\di\partial_sZ+\di[v_6+\di Y_1,Z]~.
\end{align}
As\label{Page:GenDer} we mentioned earlier, all the fields on the $\sigma^\text{th}$  slab $\mathbb{R}^{1,2}\times[p_\sigma^L, p_\sigma^R]$ are extended by zero immediately outside the slab, thus the fields have discontinuities at the ends  $p_\sigma^L$ and $p_\sigma^R$ and, potentially at any point $\lambda_j$ within the interval. In the above equations, we understand the terms $\partial_s Y_1, \partial_s Y,$ and $\partial_s Z$  as generalized derivatives. That is, at a point $\lambda_j$ the derivatives $\partial_s Y_1$ and $\partial_s Y$ contain a discontinuity term
\begin{equation} 
\delta(s-\lambda_j) \big(\lim_{r\searrow \lambda_j}Y_1(r)-\lim_{r\nearrow \lambda_j}Y_1(r)\big)~~\eand~~ \delta(s-\lambda_j) \big(\lim_{r\searrow \lambda_j}Y(r)-\lim_{r\nearrow \lambda_j}Y(r)\big)~,
\end{equation}
respectively, while $\partial_s Z$ contains terms $\delta(s-p_\sigma^L) Z(p_\sigma^L)-\delta(s-p_\sigma^R)Z(p_\sigma^R).$ This convention automatically takes into account all boundary terms appearing from any integration by parts involved in obtaining Eqs.~\eqref{Eq:Dbl}-\eqref{Eq:F2bl}.

Only four of the eight supersymmetries of CHW brane configurations are manifest in our superspace formalism. It is therefore necessary to complement the vacuum conditions $D=\bar{F}_i=\bar{G}=0$ to a system of equations that is R-symmetry equivariant, as we discuss in Section \ref{sec:Rsym}.

By the classical argument of \cite{Seiberg:1993vc, Argyres:1996eh}, the (R-symmetry completed) D- and F-flatness conditions capture the full quantum corrected Higgs branch of the moduli space of vacua in the four-dimensional gauge theory.

\subsection{Fayet-Iliopoulos terms}

Generally Fayet-Iliopoulos (FI) terms are present in the action. To preserve $SU(2)_Z$-invariance, we add FI terms for both the vector superfield $\CV$ and the chiral superfield $\CZ$:
\begin{equation}
  S_{\rm FI}=\int\dd s\, \dd^3 x \,\tr\left(\di\hat{\nu}_3(s)\int \dd^4 \theta\, \CV-\tfrac{1}{2}\hat{\nu}(s)\int \dd^2\theta\, \CZ-\tfrac{1}{2}\bar{\hat{\nu}}(s)\int \dd^2\bth\, \bar{\CZ}\right)~.
\end{equation}
These terms lead to the following contributions to the variations of the auxiliary fields:
\begin{equation}
 \begin{aligned}
\frac{\delta S_{\rm FI}}{\delta D}=&\tfrac{\di}{2}\hat{\nu}_3(s)\unit~,~~~
\frac{\delta S_{\rm FI}}{\delta F^1}=&-\tfrac{1}{2}\hat{\nu}(s)\unit~.
 \end{aligned}
\end{equation}
In the D- and F-flatness conditions \eqref{Eq:Dbl} and \eqref{Eq:F1bl}, these contributions can be absorbed by the following shift of fields (cf.\ \cite{Kapustin:1998pb}):
\begin{equation}
 Y_1\rightarrow Y_1+\tfrac{\di}{2}\unit\int^s\dd \tilde{s}\, \hat{\nu}_3(\tilde{s}) ~~\eand~~ Y\rightarrow Y+\tfrac{\di}{2}\unit\int^s\dd \tilde{s}\, \hat{\nu}(\tilde{s})~.
\end{equation}
The only effect of this shift is indeed the removal of the FI parameters in the bulk. It is only at boundaries, that they survive. Moreover, one can redefine $Y_1$ and $Y$ by a constant shift that is different on different $[p_\sigma^L,p_\sigma^R]$ intervals.  As in \cite{Cherkis:2010bn}, using these two shifts one can reduce the above FI term to the form
\begin{multline}\label{FIreduced}
  S_{\rm FI}=\sum_{\sigma=1}^k\int \dd^3 x \,\tr\left(\nu_{3\sigma}\int \dd^4 \theta\, \di\left(\CV(p_{\sigma}^L)-\CV(p_{\sigma-1}^R)\right)\right.\\
\left.  -\nu_\sigma\tfrac{1}{2}\int \dd^2\theta\, \left(\CZ(p_{\sigma}^L)-\CZ(p_{\sigma-1}^R)\right)-\bar{\nu}_\sigma\tfrac{1}{2}\int \dd^2\bth\, \left(\bar{\CZ}_1(p_{\sigma}^L)-\bar{\CZ}_1(p_{\sigma-1}^R)\right)\right)~.
\end{multline} 
The FI parameters $\nu_\sigma=\nu_{1\sigma}+\di\nu_{2\sigma}$ and $\nu_{3\sigma}$ correspond to the $\vec{\nu}_\sigma=(\nu_{1\sigma},\nu_{2\sigma},\nu_{3\sigma})$ position of the NS5-branes in $\mathbb{R}^3_Y$ in the CHW picture  of Section~\ref{Sec:CHW}.

\subsection{Defect walls with fundamental hypermultiplets }

Each fundamental impurity wall carries an $\sSU(2)_Y$ doublet $(Q_1, Q_2)$ of complex scalars together with an $\sSU(2)_Z$ doublet $(\zeta^1,\zeta^2)$ of spinors. These fields form a $\CN_{\rm 3d}=4$ hypermultiplet (cf.\ \cite{Hanany:1996ie}, \cite{DeWolfe:2001pq}) and they are localized at the value $s=\lambda_j,$ corresponding to the wall's position.  We combine them into $\CN_{\rm 3d}=2$ chiral superfields $\CQ_{1j}$ and $\CQ_{2j}$ with components $(Q_{1j},\zeta_{1j},J_{1j})$ and $(Q_{2j},\zeta_{2j},J_{2j})$ respectively.
 In doing this, we inevitably reduce the manifest R-symmetry group to the diagonal subgroup $\sSU(2)_D$ of the R-symmetry group $\sSU(2)_Z\times \sSU(2)_Y.$ The full R-symmetry is manifest once the auxiliary fields are integrated out. The superfields $\CQ_{1j}$ and $\CQ_{2j}$ are coupled to the bulk gauge field and transform under a gauge transformation $g$ in the fundamental and the antifundamental representation of the $\sU(N_\sigma)$ gauge group, respectively:
 \begin{equation}
g: \left(\begin{array}{c}
Q_{1j}\\
Q_{2j}
\end{array}\right)
\mapsto
\left(\begin{array}{c}
g^{-1}(\lambda_j)Q_{1j}\\
Q_{2j}g(\lambda_j)
\end{array}\right)~.
\end{equation}

The kinetic term is given by
\begin{equation*}\label{Fund1}
 S_{f,1}=\tfrac{1}{2}\sum_{j=1}^n\int\dd^3 x\, \dd^4\theta\,(\bar{\CQ}_{1j}\de^{-2\di \CV(\lambda_j)} \CQ_{1j}+\CQ_{2j}\de^{2\di \CV(\lambda_j)} \bar{\CQ}_{2j})~.
\end{equation*}
The full $\sSU(2)_D$-invariant Yukawa couplings in this action have to be of the form
\begin{equation}
\tfrac{1}{2}\eps_{\alpha\beta}\left(\begin{array}{c} \bar\zeta^\alpha_{1j} \\ \zeta_{2j}^\alpha\end{array}\right)^T
\left(\begin{array}{cc}Z_3 & -\bar{Z}\\ Z & -Z_3\end{array}\right)
\left(\begin{array}{c} \zeta^\beta_{1j} \\ \bar\zeta_{2j}^\beta\end{array}\right)~,
\end{equation}
and to obtain this term, we have to add the following superpotential terms:
\begin{equation*}\label{Fund2}
 S_{f,2}=\tfrac{1}{2}\sum_{j=1}^n\int \dd^3 x\left(\int \dd^2\theta\,\CQ_{2j}\CU^1(\lambda_j) \CQ_{1j}+\int \dd^2\bar{\theta}\,\bar{\CQ}_{1j}\bar{\CU}^1(\lambda_j) \bar{\CQ}_{2j}\right)~.
\end{equation*}
The contributions of the fundamental hypermultiplets to the action, $S_f=S_{f,1}+S_{f,2}$, yields the following contributions to the D- and F-flatness conditions:
\begin{equation}\label{Eq:DFf}
 \begin{aligned}
  \frac{\delta S_f}{\delta D}=&\tfrac{\di}{2}\sum_{j=1}^n(\bar{Q}_{2j}Q_{2j}- Q_{1j}\bar{Q}_{1j})\delta(s-\lambda_j)~,
  &\frac{\delta S_f}{\delta F_1}=&\tfrac{1}{2}\sum_{j=1}^nQ_{1j}Q_{2j}\delta(s-\lambda_j)~,\\
  \frac{\delta S_f}{\delta J_{1j}}=&\tfrac{1}{2}\big(\bar{J}_{1j}+Q_{2j}Z(\lambda_j)\big)~,
  &\frac{\delta S_f}{\delta J_{2j}}= &\tfrac{1}{2}\big(\bar{J}_{2j}+Z(\lambda_j)Q_{1j}\big)~.
 \end{aligned}
\end{equation}

\subsection{Defect walls with bifundamental hypermultiplets}
\label{BifundDeffects}
As mentioned above, a bifundamental impurity wall located at $p_\sigma$ cuts the space-time into two parts. This gives rise to two gauge groups $U(N_{\sigma-1})$ and $U(N_\sigma)$, acting to the left and to the right of the impurity wall, respectively.  There are two complex bifundamental  scalars $B_{1\sigma}$ and $B_{2\sigma}$ confined to the world-volume of the wall: $B_{1\sigma}$ transforms in the $(N_{\sigma-1},\bar{N}_\sigma),$ while $B_{2\sigma}$ transforms in the $(\bar{N}_{\sigma-1},N_\sigma)$ representation:
\begin{equation}
g: \left(\begin{array}{c}
B_{1\sigma}\\
B_{2\sigma}
\end{array}\right)
\mapsto
\left(\begin{array}{c}
g^{-1}(p_{\sigma-1}^R)B_{1\sigma}g(p_\sigma^L)\\
g^{-1}(p_\sigma^L)B_{2\sigma}g(p_{\sigma-1}^R)
\end{array}\right)~.
\end{equation}
They are part of a bifundamental $\CN_{\rm 3d}=4$ hypermultiplet, which we decompose into two $\CN_{\rm 3d}=2$ chiral superfields $\CB_{1\sigma}$ and $\CB_{2\sigma}$ with components $(B_{1\sigma},\xi_{1\sigma},L_{1\sigma})$ and $(B_{2\sigma},\xi_{2\sigma},L_{2\sigma})$. The complex scalars $B_{1\sigma}$ and $\bar{B}_{2\sigma}$ in this hypermultiplet again form an $\sSU(2)_Y$ doublet, cf.\ \cite{Hanany:1996ie}. 

Coupling the bifundamental superfields to the bulk gauge superfields  yields the following terms:
\begin{multline}\label{SNS51}
 S_{b,1}=\tfrac{1}{2}\sum_{\sigma=1}^k\int \dd^3 x~\tr\int \dd^4\theta\Big(\de^{2\di \CV(p_\sigma^L)}\bar{\CB}_{1\sigma}\de^{-2\di \CV(p_{\sigma-1}^R)} \CB_{1\sigma}\\
 +\de^{2\di \CV(p_{\sigma-1}^R)}\bar{\CB}_{2\sigma}\de^{-2\di \CV(p_\sigma^L)}\CB_{2\sigma}\Big)~.
\end{multline}
Again, the Yukawa couplings determine via $\sSU(2)_D$-invariance the superpotential couplings. We need to find the superfield expressions giving rise to the following terms:
\begin{equation}
\begin{aligned}
\tfrac{1}{2}\eps_{\alpha\beta}\tr&\left[
\left(\begin{smallmatrix} \bar{\xi}^\alpha_{1\sigma} & \xi_{2\sigma}^\alpha\end{smallmatrix}\right)
\left(\begin{smallmatrix}Z^R_3 & -\bar{Z}^R\\ Z^R & -Z^R_3\end{smallmatrix}\right)
\left(\begin{smallmatrix} \xi^\beta_{1\sigma} \\ \bar{\xi}_{2\sigma}^\beta\end{smallmatrix}\right)
-\left(\begin{smallmatrix} \bar{\xi}_{2\sigma}^\alpha & \xi^\alpha_{1\sigma} \end{smallmatrix}\right)
\left(\begin{smallmatrix}Z_3^L & -\bar{Z}^L\\ Z^L & -Z^L_3\end{smallmatrix}\right)
\left(\begin{smallmatrix} \xi_{2\sigma}^\beta \\ \bar{\xi}^\beta_{1\sigma}\end{smallmatrix}\right)\right]~. 
\end{aligned}
\end{equation}
This is done by adding the superpotential term
\begin{multline}\label{SNS52}
 S_{b,2}=\tfrac{1}{2}\sum_{\sigma=1}^k\int \dd^3 x~\tr\Big(\int \dd^2 \theta(\CB_{2\sigma}\CU^{1}(p_{\sigma-1}^R) \CB_{1\sigma}-\CB_{1\sigma}\CU^{1}(p_{\sigma}^L)\CB_{2\sigma})+\\
\int \dd^2 \bar{\theta}(\bar{\CB}_{1\sigma}\bar{\CU}^{1}(p_{\sigma-1}^R)\bar{\CB}_{2\sigma}-\bar{\CB}_{2\sigma}\bar{\CU}^{1}(p_{\sigma}^L)\bar{\CB}_{1\sigma})\Big)~.
\end{multline}

Finally, the positions $\vec{\nu}_\sigma$ of the NS5-branes give rise to Fayet-Iliopoulos terms at $s=p_\sigma$, cf.\  \cite{Hanany:1996ie}. As discussed above, the bulk FI-terms can be absorbed by a shift  of the scalars in the $\CN_{\rm 3d}=4$ vector multiplet. However, on the boundaries, these terms survive in the boundary contributions of the shifted scalars. These contributions, in turn, correspond to the positions of the NS5-branes. If the $\sigma$-th  NS5-brane at $s=p_\sigma$ is  positioned at $\vec{\nu}_\sigma\in\FR^3_Y,$ let $\nu=\nu_1+\di\,\nu_2.$ Then
$|\vec{\nu}_\sigma|^2=\nu_3^2+\nu\bar{\nu}~,$
and the Fayet-Iliopoulos terms are given in Eq.~\eqref{FIreduced}.
One readily checks that all terms are gauge invariant. Varying the contribution of the bifundamental matter to the gauge theory action, $S_b=S_{b,1}+S_{b,2}+S_{FI}$, with respect to the auxiliary fields yields
\begin{align}
\label{Eq:Db}
\frac{\delta S_b}{\delta  D}=&\tfrac{\di}{2}\sum_{\sigma=1}^k 
\left(\bar{B}_{1\sigma}B_{1\sigma}-B_{2\sigma}\bar{B}_{2\sigma}+\nu_{3\sigma} \right)\delta(s-p_\sigma^L)\nonumber\\
&\phantom{\tfrac{\di}{2}\sum_{\sigma=1}^k}-\left(B_{1\sigma}\bar{B}_{1\sigma}- \bar{B}_{2\sigma} B_{2\sigma}+\nu_{3\sigma}  \right)\delta(s-p_{\sigma-1}^R)~,\\
\label{Eq:Fb}
\frac{\delta S_b}{\delta   F_1}=&\tfrac{1}{2}\sum_{\sigma=1}^k\left(B_{1\sigma}B_{2\sigma}+\nu_\sigma\unit\right)\delta(s-p_{\sigma-1}^R)
-\left(B_{2\sigma}B_{1\sigma}+\nu_\sigma\unit\right)\delta(s-p_{\sigma}^L)~,\\
\frac{\delta S_f}{\delta  F_2}=&0~,\\
\frac{\delta S_b}{\delta  L_{1\sigma}}=&\tfrac{1}{2}\left(\bar{L}_{1\sigma}-Z(p_\sigma^L) B_{2\sigma}+B_{2\sigma}Z(p_{\sigma-1}^R)\right)~,\\
\frac{\delta S_b}{\delta  L_{2\sigma}}=&\tfrac{1}{2}\left(\bar{L}_{2\sigma}-B_{1\sigma}Z(p_\sigma^L)+Z(p_{\sigma-1}^R)B_{1\sigma}\right)~.&
\end{align}

\subsection{Chern-Simons boundary terms and $(p,q)$-branes}

As a side remark to the main thread of our discussion, let us briefly consider another type of defect in the CHW configuration: the $(p,q)$-branes. Their contribution to the field theory on the D3-branes is a Chern-Simons term with Chern-Simons level $k=\frac{p}{q}$, cf.\ \cite{Kitao:1998mf} and \cite{Bergman:1999na}. Intuitively speaking, the type IIB supergravity background contains an RR-scalar (axion), which gives rise to a $\theta$-term in the gauge theory on the $D3$-branes. This $\theta$-term can be turned into a Chern-Simons term at the codimension one boundary given by the $(p,q)$-brane. 

In terms of superfields, the contribution of a $(p,q)$-brane to the effective description of the CHW configuration is
\begin{equation}\label{eq:SCS}
 S_{(p,q)}=\frac{p}{8\pi q}\int \dd^3x\,\dd^4\theta\int_0^1\dd u\,\tr\left(\bar{D}^\alpha\left(\de^{-2\di \CCV(u)}D_\alpha \de^{2\di \CCV(u)}\right)\de^{-2\di \CCV(u)}\dpar_u \de^{2\di \CCV(u)}\right)~,
\end{equation}
where $\CCV(u)$ is a function on the interval $[0,1]$ satisfying the boundary conditions $\CCV(0)=0$ and $\CCV(1)=\CV$. The integral over $u$ is inserted to have the action manifestly gauge invariant, cf.\ \cite{Zupnik:1988en,Ivanov:1991fn}. The first factor under the trace is a $u$-dependent generalization $\varSigma(u)$ of the complex linear superfield $\Sigma$. In the following, we choose $\CCV(u)=u \CV$.

After going to Wess-Zumino gauge\footnote{We choose Wess-Zumino gauge for convenience. One could also define the component fields as the appropriate covariant derivatives of the vector superfield which would yield the same result.}, the action simplifies considerably and using the boundary conditions, one can perform the $u$-integration. In components, the action \eqref{eq:SCS} reads as
\begin{equation}
 \frac{\di p}{4\pi q}\tr\int \dd^3 x \big(\eps^{\mu\nu\rho}\left(A_\mu\dpar_\nu A_\rho+\tfrac{2}{3}A_\mu A_\nu A_\rho\right)+16\di \lambda^\alpha\bar{\lambda}_\alpha-2 DZ_3\big)~,
\end{equation}
and it therefore gives a new contribution to the D-flatness condition:
\begin{equation}
\frac{\delta S_{(p,q)}}{\delta D}=-\frac{\di p}{2\pi q}Z_3~.
\end{equation}
After integrating out the D-field, one obtains among others the following terms in the component action:
\begin{equation}\label{eq:components}
\int \dd^3 x\,\dd s\, \delta(s-s^{(p,q)})\tr\left(-\frac{p^2}{8\pi^2q^2}Z_3^2+\frac{\di p}{2\pi q}Z_3\nabla_sY_1+\frac{p}{4\pi q} Z_3([Z,\bar{Z}]+[Y,\bar{Y}])\right)~,
\end{equation}
where we abbreviated $\nabla_s:=\der{s}+v_6$. The last term is the same as the one obtained in a related discussion in \cite{Gaiotto:2008sa}, where an additional $\theta$-term - corresponding to our axion background - was added to a bulk theory with a supersymmetric boundary.

Assuming that the $(p,q)$-brane is oriented such that it preserves $\CN_{\rm 3d}=3$ supersymmetry, we expect the action to be invariant under the diagonal subgroup $\sSU(2)_D$ of the R-symmetry group. This leads us to add the following term to the action:
\begin{equation}
 S_{(p,q),\CN_{\rm 3d}=3}=\frac{p}{4\pi q}\int \dd^3x\,\dd s\, \delta\big(s-s^{(p,q)}\big)\left(\int\dd^2\theta \CZ^2+\int \dd^2\bar{\theta} \bar{\CZ}^2\right)~,
\end{equation}
which implies the following contribution to the $F_1$-flatness condition:
\begin{equation}
\frac{\delta S_{(p,q),\CN_{\rm 3d}=3}}{\delta F_1}=\frac{p}{4\pi q}Z~.
\end{equation}

\section{Vacuum conditions and instanton moduli spaces}
\label{Vacuum}
\subsection{D- and F-flatness conditions}

The vacuum conditions determining the Higgs branch of our gauge theory are the flatness conditions for the auxiliary fields contained in the superfields $\CV$ and $\CZ$ of the $\CN_{\rm 3d}=4$ vector superfield $\Psi$. 
Since the auxiliary fields enter the action algebraically, they can be integrated out using their equations of motion $\delta S/\delta D=0$ and $\delta S/\delta F_1=0.$ Finding $F_1$ and $D$ from these equations and substituting back into the action leads to potential terms of the forms $D^2$ and $|F_1|^2,$ here $D$ and $F_1$ stand for expressions containing derivatives that are found from the $D$ and $F_1$ equations of motion. Thus the vacuum condition is $D=0$ and $F_1=0$.  Combining Eqs.~(\ref{Eq:Dbl}, \ref{Eq:DFf}, \ref{Eq:Db}) the D-flatness condition reads\footnote{Inserting a $(p,q)$-brane at $s^{p,q}$ into our CHW configuration adds terms $-\frac{\di p}{2\pi q}Z_3\delta(s-s^{p,q})$ and $\frac{p}{4\pi q}\delta(s-s^{p,q})Z$ to the left hand side of Eqs.~\eqref{BPSD} and \eqref{BPSF1} respectively.}
\begin{multline}\label{BPSD}
 \nabla_s Y_1+\tfrac{\di}{2}[Z,\bar{Z}]+\tfrac{\di}{2}[Y,\bar{Y}]+\tfrac{\di}{2}\sum_{j=1}^n(Q_{1j}\bar{Q}_{1j}-\bar{Q}_{2j}Q_{2j}) \delta(s-\lambda_j) \\
+\tfrac{\di}{2}\sum_{\sigma=1}^k\left(B_{2\sigma}\bar{B}_{2\sigma}-\bar{B}_{1 \sigma}B_{1\sigma}-\nu_{3\sigma}\unit\right)\delta(s-p_{\sigma}^L)\\
+\left(B_{1\sigma}\bar{B}_{1\sigma}-\bar{B}_{2\sigma}B_{2\sigma}+\nu_{3\sigma}\unit\right)\delta(s-p_{\sigma-1}^R)=0~,
\end{multline}
and Eqs.~(\ref{Eq:F1bl}, \ref{Eq:DFf}, \ref{Eq:Fb}) lead to the F-flatness condition
\begin{multline}\label{BPSF1}
\nabla_sY+\di[Y_1,Y]+\tfrac{\di}{2}\sum_{j=1}^nQ_{1j}Q_{2j} \delta(s-\lambda_j) \\
+\tfrac{\di}{2}\sum_{\sigma=1}^k\left(B_{1\sigma}B_{2\sigma}+\nu_\sigma\unit\right)\delta(s-p_{\sigma-1}^R)
-\left(B_{2\sigma}B_{1\sigma}+\nu_\sigma\unit\right)\delta(s-p_{\sigma}^L)=0~,
\end{multline}
where again $\nabla_s:=\der{s}+v_6$.

The G- and F$_2$-flatness conditions from the superfields $\CY$ and $\CX$ contained in the $\CN_{\rm 3d}=4$ hypermultiplet $\tilde{\Upsilon}$ are $G=F_2=0$, which amount to
\begin{equation}\label{BPSF2}
 \begin{aligned}
\nabla_sZ+\di[Y_1,Z]=0~,
 \end{aligned}
\end{equation}
and
\begin{equation}\label{BPSG}
 \begin{aligned}
{}[Z,Y]=0~.
 \end{aligned}
\end{equation}

At a position $\lambda_j$, where a fundamental impurity wall is located, we have the following additional  equations for the auxiliary fields of $\CQ_{1j}$ and $\CQ_{2j}$:
\begin{equation}\label{BPSJi}
 \begin{aligned}
  \bar{J}_{1j}+Q_{2j}Z(\lambda_j)&=0~,~~~
  \bar{J}_{2j}+Z(\lambda_j)Q_{1j}&=0~.
 \end{aligned}
\end{equation}

A bifundamental wall at $p_\sigma$ gives rise to two flatness conditions on its two sides arising from the auxiliary fields of $\CB_{1\sigma}$ and $\CB_{2\sigma}$:
\begin{equation}\label{BPSLi}
 \begin{aligned}
\bar{L}_{1\sigma}-Z(p_\sigma^L) B_{2\sigma}+B_{2\sigma}Z(p_{\sigma-1}^R)=0~,\\
 \bar{L}_{2\sigma}-B_{1\sigma}Z(p_\sigma^L)+Z(p_{\sigma-1}^R)B_{1\sigma}=0~.
 \end{aligned}
\end{equation}

\subsection{R-symmetry consequences}\label{sec:Rsym}

The equations derived above yield vacua preserving the $\CN_{\rm 3d}=2$ supersymmetries which are manifest in our superspace formulation. These supersymmetries are a subset of the $\CN_{\rm 3d}=4$ supersymmetries actually preserved by CHW configurations. To obtain the equations describing the vacua of our $\CN_{\rm 3d}=4$ supersymmetric theory, one needs to consider the set of flatness conditions obtained by all possible rewritings of  the $\CN_{\rm 3d}=4$ supersymmetries in $\CN_{\rm 3d}=2$ language. This can be achieved by complementing the above equations such that the new set of equations is invariant under the R-symmetry group $\sSU(2)_Z$ and is equivariant under $\sSU(2)_Y$.

First, equations \eqref{BPSF2} and \eqref{BPSG} have to be replaced by
\begin{equation}\label{eq:M1}
  {}[Z,Y]=[Z_3,Y]=[Z,Y_1]=[Z_3,Y_1]=0~~\eand~~\nabla_sZ=\nabla_sZ_3=0~,
\end{equation}
and thus $Z$ and $Z_3$ ought to be covariantly constant in the bulk along the $s$-direction. On a branch with nontrivial vacuum expectation values of $Y_1(s), Y_2(s),$ and $Y_3(s),$ i.e. on the Higgs branch,  Eqs.~\eqref{eq:M1} imply that $Z$ and $Z_3$ vanish\footnote{$Z$ and $Z_3$ vanish unless we consider the mixed branch with all of the Higgs fields $Z, Z_3, Y,$ and $Y_3$ commuting with each other.}.  Correspondingly, \eqref{BPSD} and \eqref{BPSF1} become
\begin{multline}\label{eq:M2}
 \nabla_s Y_1+\tfrac{\di}{2}[Y,\bar{Y}]+\tfrac{\di}{2}\sum_{j=1}^n(Q_{1j}\bar{Q}_{1j}-\bar{Q}_{2j}Q_{2j}) \delta(s-\lambda_j) \\
+\tfrac{\di}{2}\sum_{\sigma=1}^k\left(B_{2\sigma}\bar{B}_{2\sigma}-\bar{B}_{1 \sigma}B_{1\sigma}-\nu_{3\sigma}\unit\right)\delta(s-p_{\sigma}^L)\\
+\left(B_{1\sigma}\bar{B}_{1\sigma}-\bar{B}_{2\sigma}B_{2\sigma}+\nu_{3\sigma}\unit\right)\delta(s-p_{\sigma-1}^R)=0~,
\end{multline}
\begin{multline}\label{eq:M3}
\nabla_sY+\di[Y_1,Y]+\tfrac{\di}{2}\sum_{j=1}^nQ_{1j}Q_{2j} \delta(s-\lambda_j) \\
+\tfrac{\di}{2}\sum_{\sigma=1}^k\left(B_{1\sigma}B_{2\sigma}+\nu_\sigma\unit\right)\delta(s-p_{\sigma-1}^R)
-\left(B_{2\sigma}B_{1\sigma}+\nu_\sigma\unit\right)\delta(s-p_{\sigma}^L)=0~.
\end{multline}
The R-symmetry completions of equations \eqref{BPSJi} and \eqref{BPSLi} are simple to write down, however, they will not be relevant for  our discussion.

\subsection{Masses}

If a Z-branch were present, the D3-branes would be able to break up on the NS5-brane. This would give masses to the bifundamentals located at the intersection of the stack of D3-branes with the NS5-brane which would be  proportional to the distance between the endpoints of the broken D3-branes. For diagonal $Z_3$ and $Z$ the diagonal components are $Z_{3aa}$ and $Z_{aa},$ and, as in the case of the D5-branes, this distance is split into the complex $Z^{L}_{aa}-Z^{R}_{bb}$ and real $Z_{3aa}^L-Z_{3bb}^R$ components. The mass contribution of the former appears directly in the action \eqref{SNS51}, while the contribution of the latter arises after integrating out the auxiliary fields $L_{i\sigma}$. Altogether, we have the expected result for the mass of the bifundamental hypermultiplets:
\begin{equation}
 m^2_{B}=\left|Z^{L}_{aa}-Z^{R}_{bb}\right|^2+\left(Z_{3aa}^L-Z_{3bb}^R\right)^2~.
\end{equation}

Mass terms for the fundamental hypermultiplet located at D5-brane positions arise from a finite distance between the D3-branes and the D5-branes in $\FR^3_Z$. Although our D5-branes will be located at $\vec{z}=\vec{0}$, let us briefly comment on the more general situation of a  D5-brane at $s=\lambda_j$ and $\vec{z}^{D5}=(z^{D5},z^{D5}_3)\neq \vec{0}$. This situation is described by the following gauge theory action:
\begin{equation}\label{FundMass}
 \begin{aligned}
 S_{f,1}&=\tfrac{1}{2}\int \dd s\,\dd^3 x\int \dd^4\theta\,(\bar{\CQ}_{1j}\de^{-2\di (\CV-\di\theta\bar\theta z_{3,j}^{D5}\unit)} \CQ_{1j}+\CQ_{2j}\de^{2\di (\CV-\di\theta\bar\theta z_{3,j}^{D5}\unit)} \bar{\CQ}_{2j})\delta(s-\lambda_j)~,\\
 S_{f,2}&=\tfrac{1}{2}\int \dd s\,\dd^3 x\left(\int \dd^2\theta\,\CQ_{2j}(\CU^1-z^{D5}_j)\CQ_{1j}+c.c.)\right)\delta(s-\lambda_j)~.
 \end{aligned}
\end{equation}
The modified terms preserve both supersymmetry and gauge invariance as they can be viewed as constant shifts of scalar fields. 

The eigenvalues of $Z$ and $Z_3$ correspond to the positions of the various D3-branes in the stack. The distances of the D3-branes to the D5-branes at $\vec{z}^{D5}$ are then given by the eigenvalues of the matrices $Z-z^{D5}\unit$ and $Z_3-z_3^{D5}\di\unit$. After integrating out the auxiliary fields $J_{1,2j}$ and diagonalizing $Z$ and $Z_3$, one obtains the following mass terms for the fundamental hypermultiplets:
\begin{equation}
 m^2_{\CQ_p}=\left|Z_{aa}-z^{D5}_p\right|^2+\left(Z_{3aa}-z^{D5}_{3,p}\di\unit\right)^2~,
\end{equation}
which is the expected result.

\subsection{Interpretation of the Y-branch equations}

The moduli space equations \eqref{BPSD}, \eqref{BPSF1}, \eqref{BPSF2} and \eqref{BPSG} reflect many of the interesting phenomena in CHW configurations. Besides the generation of masses from moving the various five-branes as discussed above, the breaking of D3-branes on both the NS5-branes and the D5-branes can be seen in the gauge theory. In the following, however, we are interested in using these equations to describe instantons on multi Taub-NUT spaces.  In particular, we are about to demonstrate that the moduli space of the latter is the Y-branch of the gauge theory we are studying here.

Yang-Mills instantons on multi Taub-NUT spaces can be described in terms of bows \cite{Cherkis:2010bn}.  Bows generalize quivers, and as for a quiver, one can define representations of a bow.  Detailed explanations of these and other terms related to bows can be found in \cite{Cherkis:2010bn}. Each bow representation can be viewed in two ways: as a description of instantons of given charges on multi Taub-NUT spaces or as defining a gauge theory with impurities such as considered above. In the latter interpretation, each edge in the bow corresponds to an impurity wall with a bifundamental multiplet, while each marked point corresponds to an impurity wall with a fundamental multiplet on it.  The representation ranks determine the ranks of  the unitary gauge groups in the bulk between neighboring impurity walls. 

A representation $\mathfrak{R}$ of a bow determines an affine space ${\rm Dat}(\mathfrak{R})$, the bow data.  In the gauge theory, ${\rm Dat}(\mathfrak{R})$ can be thought of as the configuration space of the scalar fields in the chiral superfields which parameterize the Y-branch. Because of $\CN_{\rm 3d}=4$ supersymmetry, the space ${\rm Dat}(\mathfrak{R})$ is hyperk\"ahler, and there is a natural action of a gauge group $\CG$ on ${\rm Dat}(\mathfrak{R})$ which preserves the hyperk\"ahler structure. Therefore, one can construct the hyperk\"ahler quotient $\CM={\rm Dat}(\mathfrak{R})/\!/\!/\CG$. The corresponding moment map was given in \cite{Cherkis:2010bn}.  The gauge theory counterpart of the hyperk\"ahler quotient reduction  amounts to imposing the D- and F-flatness conditions and dividing by the action of the gauge group.  

In their complex form, the bow equations of \cite{Cherkis:2010bn}, which are the moment map conditions,  exactly agree with the vacuum equations \eqref{eq:M2} and \eqref{eq:M3}, which define the Y-branch of the space of vacua of our gauge theory.  We thus conclude that the Y-branch coincides with the moduli space of instantons on a multi-Taub-NUT space.  The dictionary establishing the correspondence of the quantities of the gauge theory with impurities with those of an instanton on the multi-Taub-NUT space is given in Table~\ref{Dictionary} below.  This correspondence can be established either via T-duality or using the transform of \cite{Cherkis:2010bn}.  T-duality relates a CHW configuration to a configuration of D6-branes wrapping multi-Taub-NUT space with D2-branes within their world-volumes.  At low energies, the latter brane configuration is described by Yang-Mills instantons on the wrapped space.  Here, we obtained an independent gauge theoretic verification of this correspondence.  In the process, we also gained some insight of how one might approach the other branches.  We also have a gauge theoretic interpretation of the bow reciprocity, which manifests itself as electric-magnetic duality of the gauge theory with impurities.  We are about to use electric-magnetic duality to extract the asymptotic of the Y-branch in the directions in which the  gauge group is maximally broken.
\begin{table}[htdp]
\begin{center}
\begin{tabular}{p{190pt} | p{190pt}}
\multicolumn{1}{c}{Gauge theory with impurity walls} & \multicolumn{1}{|c}{Instantons on  multi-Taub-NUT}\\
\hline
\hline
Gauge group ranks &  {Instanton number + monopole charges}\\
\hline
Number of fundamental walls, $n$ & Rank of the structure group, $U(n)$\\
\hline
Number of bifundamental walls, $k$ & Number of Taub-NUT centers, $k$\\
\hline
Periodicity of the transverse coordinate $s$ & Taub-NUT mass parameter\\
\hline
Positions of the fundamental walls & Conjugacy class of the holonomy 
around the TN circle at infinity\\
\hline
Positions of the bifundamental walls & Self-dual noncommutativity parameters of the multi-Taub-NUT\\
\hline
Fayet-Iliopoulos parameters & Positions of the Taub-NUT centers\\
\end{tabular}
\end{center}
\caption{Correspondence between the gauge theory and instanton parameters.}
\label{Dictionary}
\end{table}%

\section{Asymptotic of the Y-branch}\label{Sec:OneLoop}

Until this moment, we identified the Y-branch with the Higgs branch of the gauge theory with impurity walls.  The advantage of this consideration was that it produced an exact description of the metric on the Y-branch \cite{Seiberg:1993vc, Argyres:1996eh} in terms of Eqs.~\eqref{eq:M1} and \eqref{eq:M2}.  If one is interested in the asymptotic behavior of these metrics, one can either use twistorial techniques developed by Bielawski \cite{Bielawski:1997, Bielawski:1998hk, Bielawski:1998hj} or apply the monopole dynamics techniques of Manton and Gibbons \cite{Manton:1985hs, Gibbons:1995yw}.  Here we have yet another approach, which is entirely in the domain of the quantum gauge theory.  
	
A different description of the Y-branch emerges after applying electric-magnetic duality to the gauge theory we considered so far.  If the original gauge theory had $n$ fundamental impurity walls positioned at $s=\lambda_j$ and $k$ bifundamental impurity walls at $s=p_\sigma,$ then the dual gauge theory has $k$ fundamental impurity walls at $s=p_\sigma$ and $n$ bifundamental impurity walls positioned at $s=\lambda_j.$  This is one of the simplest impurity wall dualities.  A much more general situation is studied in \cite{Gaiotto:2008ak}.  The Y-branch is the Coulomb branch of the latter theory and the metric on it receives both perturbative and nonperturbative corrections.  At a generic point on the Coulomb branch, the gauge symmetry is maximally broken.  The metric on the Y-branch is given by quantum corrected gauge couplings of the surviving gauge group.  Eigenvalues of the $Y_1, Y_2$ and $Y_3$ Higgs fields provide good asymptotic coordinates on the Y-branch.  We are interested in finding the metric in the asymptotic directions in which the difference of any two eigenvalues of $Y_j$ becomes large\footnote{To be exact, it suffices that the absolute value of the difference of any two vectors $\vec{y}_a=(y^{1,a}, y^{2,a}, y^{3,a})$ of eigenvalues is large.  These eigenvalues are defined in the next section.}.  Via a one-loop computation we obtain the leading metric behavior.
	
It is worth emphasizing at this point that the electric-magnetic duality interchanges not only the two types of impurity walls, but also the roles of the $Y$ and $Z$ Higgs fields.  Thus the electric-magnetic dual theory is given by 
\begin{itemize}
\renewcommand{\labelitemi}{$\cdot$}
\item the bulk action \eqref{actionbulk} (without the FI terms \eqref{FIreduced}), with the $Y^j$ and $Z^j$ fields interchanged, 
\item fundamental multiplet contributions \eqref{FundMass}, with $p_\sigma$ in place of $\lambda_j,$ $\CY$ in place of $\CU^1,$ $z_\sigma^\text{D5}=\nu_\sigma$ and $z_{3,\sigma}^\text{D5}=\nu_{3,\sigma},$ and 
\item bifundamental multiplet contributions (\ref{SNS51}, \ref{SNS52}) with $\lambda_{j-1}^R$ in place of $p_{\sigma-1}^R,$ $\lambda_{j}^L$ in place of $p_{\sigma}^L,$ and $\CY$ in place of $\CU^1.$ 
\end{itemize}
The resulting flatness conditions, such as \eqref{eq:M1} or \eqref{eq:M2}, are given by the same expressions but with letters $Y$ and $Z$ interchanged.  

On the Coulomb branch, the $G$ and $F_2$ flatness conditions augmented by R-symmetry imply that the nonvanishing Higgs fields $Y$ and $Y_1$  parameterizing the Coulomb branch are covariantly constant in $s,$ see Eq.~\eqref{eq:M1}.  As a result, in the extreme infrared the theory is effectively three-dimensional and we can perform our one-loop computation in a three dimensional theory with gauge group $\displaystyle\mathop{\times}_{j=1}^n U(R_j).$  The gauge coupling of the component $U(R_j)$ is $g_{\rm 4d}/\sqrt{\lambda_{j+1}-\lambda_j},$ so that 
\begin{equation}
\frac{1}{g^2_{3d, j}}=\frac{\lambda_{j+1}-\lambda_j}{g^2_{\rm 4d}}~.
\end{equation}
The ultraviolet spectrum of this theory is comprised of bifundamental supermultiplets in the $(N_{j-1}, \bar{N}_{j})$ and $(\bar{N}_{j-1},  N_j)$ representations and some fundamental multiplets.  The number of the fundamental multiplets in $N_j$ and $\bar{N}_j$ of $U(N_j)$ equals the number of points $p_\sigma$ between $\lambda_j$ and $\lambda_{j+1}.$  The mass of the fundamental multiplet associated to the point $p_\sigma$ equals $\vec{\nu}_\sigma.$

The computation itself uses the background field method quite literally as discussed in \cite{Peskin:1995ev}, section 16.6. This background field calculation was done for pure $\CN_{\rm 3d}=4$ super Yang-Mills theory with gauge group $\sSU(2)$ in \cite{Dorey:1997ij} and it was extended to fundamental matter in \cite{Dorey:1998kq}. For our discussion, however, we need the corresponding result for arbitrary gauge group and both fundamental and bifundamental matter, which we derive in some detail below.

\subsection{The Coulomb branch}

It is sufficient to focus on $\CN_{\rm 3d}=4$ euclidean super Yang-Mills theory with gauge group $\sU(N_L)\times \sU(N_R)$. The field content consists of an $\CN_{\rm 3d}=4$ vector multiplet containing a gauge potential $A_\mu=A_\mu^L+A_\mu^R$, two 3d Majorana spinors $\lambda^{L,R}$ and $\chi^{L,R}$ and three real Higgs fields $Y^i_{L,R}$. We also allow for hypermultiplets in the bifundamental, the adjoint and the fundamental representations of the gauge group. Their component fields are labeled by $h_k$, $k=1,2$ for the complex scalars and $\psi_k$ for the spinors of $\sSO(1,2)$. The relevant kinetic terms in the action are given by 
\begin{equation}
\begin{aligned}
 S^{L,R}_{\rm kin,~gauge}&=\frac{1}{g^2_{3d\, L,R}}\int \dd^3 x~ \tr\Big(\tfrac{1}{4} F^{L,R}_{\mu\nu}F_{L,R}^{\mu\nu}+\tfrac{1}{2}\nabla^{L,R}_\mu Y^{L,R}_i\nabla^{L,R}_\mu Y^{L,R}_i\\
&\hspace{3.5cm}+\di \bar{\lambda}^{L,R}\nablas^{L,R}\lambda^{L,R}+\di \bar{\chi}^{L,R}\nablas^{L,R}\chi^{L,R}\Big)~,\\
 S^{L,R}_{\rm kin,~hyper}&=\frac{1}{g^2_{3d\, L,R}}\int \dd^3 x~\tr\left(\nabla_\mu h^\dagger_k\nabla^\mu h_k+\di\bar{\vartheta}_k\nablas\vartheta_k\right)~.
\end{aligned}
\end{equation}
Here, $\nablas:=\bar{\sigma}^\mu\nabla_\mu$. The covariant derivatives of  the hypermultiplets are determined by their representation. We did not write down any potential terms, as we shall not need them.

As generators for $U(N)$, we  use antihermitian linear combinations of the matrices $(\tau_{ab})_{ij}=\delta_{ai}\delta_{bj}$, $a,b=1,\ldots,N$, which satisfy the normalization condition $\tr(\tau_{ab}^T\tau_{ab})=1$. In the latter equation, there is no sum implied. For indices of the kind of $a,b$, we  always make the sums explicit in all formulas. The Cartan subalgebra of the gauge group is generated by the elements $\tau_{aa}$. Note that $\sum_a \tau_{aa}$ gives the $\au(1)$-part of $\au(N)$, while the linear combinations of generators $\tau_{aa}-\tau_{bb}$ for $a\neq b$ span the Cartan subalgebra of $\asu(N)\subset \au(N)$. For the generators $\tau_{ab}$, the following list of identities holds:
\begin{equation}
\begin{aligned}
\tau_{ab}\tau_{cd}=\delta_{bc}\tau_{ad}~,~~~{\rm ad}_{\tau_{ab}}(\tau_{cd})=&[\tau_{ab},\tau_{cd}]=\tau_{ad}\delta_{bc}-\tau_{cb}\delta_{da}~,\\ [\tau_{aa},\tau_{bb}]=0~,~~~&[{\rm ad}_{\tau_{aa}},{\rm ad}_{\tau_{bb}}]=0~. 
\end{aligned}
\end{equation}
To distinguish the generators of $\sU(N_L)$ from those of $\sU(N_R)$, we  write $\tau_{ab}^L$ and $\tau_{ab}^R$, where necessary.

We are interested in a generic point on the Coulomb branch of the theory, where the gauge groups $\sU(N_L)\times \sU(N_R)$ are maximally Higgsed to $\sU(1)^{N_L}\times \sU(1)^{N_R}$ due to the Higgs scalars in the vector multiplet acquiring a generic vacuum expectation value (vev). As the gauge group is abelian, we can dualize the $N_L+N_R$ resulting photons into periodic scalars $\sigma^a_L\sim\sigma^a_L+1$, $a=1,\ldots,N_L$, and $\sigma^a_R\sim\sigma^a_R+1$, $a=1,\ldots,N_R$, parameterizing $T^{N_L+N_R}$. 
More explicitly, we add the following surface term to the action:
\begin{equation}
 S_{\rm \theta}=\frac{\di}{8\pi}\int \dd^3 x~\left(\eps^{\mu\nu\rho} \sum_{a=1}^{N_L}\sigma^a_L\dpar_\mu F^{L,a}_{\nu\rho}+\eps^{\mu\nu\rho} \sum_{a=1}^{N_R}\sigma^a_R\dpar_\mu F^{R,a}_{\nu\rho}\right)~,
\end{equation}
which, after integrating out the abelian field strength, yields the kinetic term for the dual photons $\sigma^a$:
\begin{equation}
 S_{\rm kin,~dual}=\frac{4 g_{3d\, L}^2}{(8\pi)^2}\int \dd^3x~\sum_{a=1}^{N_L}\tfrac{1}{2}\dpar_\mu\sigma^a_L\dpar^\mu\sigma^a_L+
 \frac{4 g_{3d\, R}^2}{(8\pi)^2}\int \dd^3x~\sum_{a=1}^{N_R}\tfrac{1}{2}\dpar_\mu\sigma^a_R\dpar^\mu\sigma^a_R~.
\end{equation}
The open part of the classical moduli space $\CM_{cl}$ of this theory is a  subspace of the direct product of the space $\FR^{3(N_L+N_R)}$ parameterized by mutually commuting scalar fields $Y^i=\sum_{a=1}^{N_L+N_R} y^{i,a}\tau_{aa}$ times the torus parameterized by the vev of the dual photons $T^{N_L+N_R}$. At the classical level, gauge symmetry is enhanced at the set $\Delta$ of points in $\FR^{3(N_L+N_R)}$ with $y^{i,a}=y^{i,b}$ for any $a$ and $b\neq a$. We remove this set from $\FR^{3(N_L+N_R)}$. The resulting space still has to be factored by the symmetric group $S_{N_L}\times S_{N_R}$ to eliminate permutations of the eigenvalues $y^{i,a}$. Away from $\Delta$, the classical moduli space thus has the form
\begin{equation}
\frac{(\FR^{3(N_L+N_R)}\backslash \Delta)\times T^{N_L}\times T^{N_R}}{S_{N_L}\times S_{N_R}}~,
\end{equation}
with its flat metric.  In the next section, we compute the one-loop corrections to the K{\"a}hler metric on this moduli space. 

\subsection{One-loop correction to the gauge couplings}
\label{Sec:Corr}

To perform the one-loop background field computation, we split the Yang-Mills fields into a low-momentum background component\footnote{The background component is supersymmetric and satisfies the equations of motion.} and a high-momentum part and then integrate out the latter.  Explicitly, we rewrite the Yang-Mills action using this splitting and keep only terms up to second order in the high-momentum parts. The functional integrals over the high-momentum fields are Gau{\ss}ian and can be trivially performed. We can read off their contributions to the effective action from the logarithm of the product of the resulting determinants.  

In terms of ordinary perturbation theory, this means that we compute Feynman diagrams with one loop which have low-momentum parts as external legs and high-momentum parts in the loop. We then replace these diagrams by effective vertices.

As the gauge group is maximally broken, we have the vacuum expectation values  of $Y^i$ aligned in the direction of the Cartan generators: $Y^i=\sum_a y^{i,a}\tau_{aa}$. Furthermore, we combine the gauge field $A_\mu$ and the three scalars $Y^i$ into a six-dimensional gauge field $\CA_M$, $M=0,\ldots,5$ so that $\CA_\mu=A_\mu$ and $\CA_{i+2}=Y^i$, cf.\ \cite{Dorey:1997ij}. Similarly, we combine the two Majorana spinors of $\sSO(1,2)$, $\lambda$ and $\chi$, into a Weyl spinor $\eta$ of $\sSO(1,5)$. 

As usual in the background field method, we now split our fields $\varphi$ into slowly oscillating background fields $\mathring{\varphi}$ and a high-momentum part\footnote{For simplicity, we drop the labels $L,R$ when no confusion can arise.} $\tilde{\varphi}$:
\begin{equation}
 \begin{aligned}
  \CA_M&=\left\{\begin{array}{ll}\mathring{\CA}_M+\tilde{\CA}_M~&,~~~M\leq 2~,\\
  \mathring{\CA}_M+\sum_a y^{M-2,a}\tau_{aa}+\tilde{\CA}_M~&,~~~M\geq 3~,                
\end{array}\right. \\
  \eta&=\mathring{\eta}+\tilde{\eta}~,~~~h_i=\mathring{h}_i+\tilde{h}_i~,~~~\vartheta_i=\mathring{\vartheta}_i+\tilde{\vartheta}_i~.
 \end{aligned}
\end{equation}
The low-momentum background fields $\mathring{\CA}$, $\mathring{\eta}$, $\mathring{h}_i$ and $\mathring{\vartheta}_i$ are aligned in the direction of the Cartan generators. 

To integrate out the high-momentum fields, we plug this expansion into the action. We gauge fix the action and introduce ghosts.  This is done completely analogously to \cite{Dorey:1997ij}. There is thus an additional complex triplet of ghosts. The ghosts are not split into low- and high-momentum parts. 

One can drop linear terms in any of the high-momentum fields, as they multiply terms proportional to equations of motion of the background fields. We also drop terms of higher order than two, as these do not contribute to the renormalization of purely low-momentum vertices at one loop. The key observation is that the remaining terms in the action are all of the form
\begin{equation}\label{eq:actionStandardForm}
 \int \dd^3 x\,\tr\left(\tilde{\varphi}^\dagger\left(\Delta_{\varphi}^{\kappa_\varphi}\right)\tilde{\varphi}\right)~,
\end{equation}
where $\varphi$ denotes an arbitrary field and the power $\kappa_\varphi=1$ for bosons and $\kappa_\varphi=\frac{1}{2}$ for spinor fields. The ghost contribution to the action is also of this form with $\kappa_{\rm gh}=1$.

The resulting functional integrals are Gau{\ss}ian and can be easily performed: They lead to determinants of the $\Delta_\varphi$ raised to a certain power. Rewriting them as exponentials to read off the one-loop corrections to the action yields
\begin{equation}
 \delta S_{1\ell}=\sum_\varphi \pi_\varphi \tr(\log \Delta_\varphi)~,~~~ \pi_{\CA}=-\tfrac{1}{2}~,~~ \pi_{\eta}=\tfrac{1}{2}~,~~ \pi_{\rm gh}=1~,~~ \pi_{h}=-1~,~~ \pi_{\vartheta}=\tfrac{1}{2}~,
\end{equation}
where the $\pi_\varphi$ are the powers of the determinants appearing from the Gau{\ss}ian functional integral. The trace symbol here denotes a trace over gauge and spinor indices as well as all necessarily implied integrals over momentum spaces. Summarizing, we perform the following approximation of the functional integral over the higher momentum modes:
\begin{equation}
 \begin{aligned}
  \CZ&=\int\left(\prod_{\varphi}\CCD\varphi\right)\de^{-\frac{\di}{\hbar}S[\varphi]}\int\left(\prod_{\tilde{\varphi}}\CCD\tilde{\varphi}\right)\de^{-\frac{\di}{\hbar}\sum_{\tilde{\varphi}}\int \dd^3 x\,\tr\left(\tilde{\varphi}^\dagger\left(\Delta_{\varphi}^{\kappa_\varphi}\right)\tilde{\varphi}\right)}\\
&\approx\int\left(\prod_{\varphi}\CCD\varphi\right)\de^{-\frac{\di}{\hbar}\left(S[\varphi]+\sum_\varphi \pi_\varphi \tr(\log \Delta_\varphi)\right)}~.
 \end{aligned}
\end{equation}
In order to remain consistent with our one-loop approximation scheme, we should compute $\delta S_{1\ell}$ only up to second order in the high-momentum fields. This is what we do below.

The kernels $\Delta_\varphi$ depend exclusively on the spin of the fields $\varphi$ and their gauge representation. Its explicit form is easily obtained by adapting the formulas in \cite{Dorey:1997ij} or \cite{Peskin:1995ev}:
\begin{equation}\label{eq:Deltas1}
 \begin{aligned}
  \Delta_\varphi&=-\dpar^2+\Delta^{(1)}_\varphi+\Delta^{(2)}_\varphi+\Delta^{(J)}_\varphi~,\\
 \Delta^{(1)}_\varphi&=\di\left\{\dpar^\mu,\sum_{a=1}^{N_L}\mathring{\CA}_\mu^{L,a} \tau^L_{aa}+\sum_{a=1}^{N_R}\mathring{\CA}_\mu^{R,a} \tau^R_{aa}\right\}~,\\
 \Delta^{(2)}_\varphi&=\sum_{a,b,L/R}\left(\mathring{\CA}^{L/R,a}_M\mathring{\CA}^{L/R,b}_M- \sum_{i=1}^3 2y^{i,a}_{L/R}\mathring{\CA}^{L/R,b}_{i+2}-y^{i,a}_{L/R}y^{i,b}_{L/R}\right)\tau^{L/R}_{aa}\tau^{L/R}_{bb}~,\\
 \Delta^{(J)}_\varphi&=\sum_{a=1}^{N_L}\mathring{\CF}_{MN}^{L,a}\tau^L_{aa} J^{MN}_\varphi+\sum_{a=1}^{N_R}\mathring{\CF}_{MN}^{R,a}\tau^R_{aa} J^{MN}_\varphi~.
 \end{aligned}
\end{equation}
Here, the sum in $\Delta^{(2)}_\varphi$ runs over all possible combinations of indices $a,b$ and gauge potentials and Cartan generators of $U(N_L)$ and $U(N_R)$. The generators $J^{MN}_\varphi$ are the generators of the Lorentz group in six dimensional Minkowski space in the representation of  the field $\varphi$. For fields $\varphi_{\rm f,L}$, $\varphi_{\rm ad,L}$ and $\varphi_{\rm bf,LR}$, in the fundamental representation of $\sU(N_L)$, the adjoint representation of $\sU(N_L)$ and the bifundamental representation of $\sU(N_L)\times \sU(N_R)$, we have
\begin{equation}\label{eq:Deltas2}
 \begin{aligned}
    \Delta^{(1)}_\varphi\varphi_{\rm f,L}&=\di\left\{\dpar^\mu,\sum_{a=1}^{N_L}\mathring{\CA}_\mu^{L,a} \tau^L_{aa}\right\}\varphi_{\rm f,L}~,\\
 \Delta^{(2)}_\varphi\varphi_{\rm ad,L}&=\sum_{a,b=1}^{N_L}\left(\mathring{\CA}^{L,a}_M\mathring{\CA}^{L,b}_M- \sum_{i=1}^3 2y^{i,a}_{L}\mathring{\CA}^{L,b}_{i+2}-y^{i,a}_{L}y^{i,b}_{L}\right)[\tau^{L}_{aa},[\tau^{L}_{bb},\varphi_{\rm ad,L}]]~,\\
 \Delta^{(2)}_\varphi\varphi_{\rm bf,LR}&=\sum_{a,b=1}^{N_L}\left(\mathring{\CA}^{L,a}_M\mathring{\CA}^{L,b}_M- \sum_{i=1}^3 2y^{i,a}_{L}\mathring{\CA}^{L,b}_{i+2}-y^{i,a}_{L}y^{i,b}_{L}\right)\tau^{L}_{aa}\tau^{L}_{bb}\varphi_{\rm bf,LR}\\
&~~-\sum_{a=1}^{N_L}\sum_{b=1}^{N_R}\left(\mathring{\CA}^{L,a}_M\mathring{\CA}^{R,b}_M-\sum_{i=1}^32 y^{i,a}_{L}\mathring{\CA}^{R,b}_{i+2}-y^{i,a}_{L}y^{i,b}_{R}\right)\tau^{L}_{aa}\varphi_{\rm bf,LR}\tau^{R}_{bb}\\
&~~-\sum_{a=1}^{N_R}\sum_{b=1}^{N_L}\left(\mathring{\CA}^{R,a}_M\mathring{\CA}^{L,b}_M-\sum_{i=1}^32 y^{i,a}_{R}\mathring{\CA}^{L,b}_{i+2}-y^{i,a}_{R}y^{i,b}_{L}\right)\tau^{L}_{bb}\varphi_{\rm bf,LR}\tau^{R}_{aa}\\
&~~+\sum_{a,b=1}^{N_R}\left(\mathring{\CA}^{R,a}_M\mathring{\CA}^{R,b}_M-\sum_{i=1}^32 \vec{y}\,^{a}_{R}\mathring{\CA}^{R,b}_5-y^{i,a}_{R}y^{i,b}_{R}\right)\varphi_{\rm bf,LR}\tau^{R}_{aa}\tau^{R}_{bb}~.
 \end{aligned}
\end{equation}
We also recall that all background fields are in the Cartan subalgebra and thus 
\begin{equation}
 \mathring{\CF}^a_{MN}:=\dpar_M\mathring{A}^a_{N}-\dpar_N\mathring{A}^a_{M}~.
\end{equation}

To evaluate the effective action $\delta S_{1\ell}$ to quadratic order in the background fields, we Taylor expand the logarithm around
\begin{equation}\label{def:propagator}
-\dpar^2-\sum_{i,a,b,L/R}y^{i,a}_{L/R}y^{i,b}_{L/R}\tau^{L/R}_{aa}\tau^{L/R}_{bb}=:\CG^{-1}~,
\end{equation}
the inverse of the propagator $\CG$, and drop terms of higher than quadratic order in $\mathring{\CA}_M$. Up to an irrelevant constant, we obtain:
\begin{equation}\label{eq:1L-1}
 \delta S_{1\ell}=\sum_\varphi \pi_\varphi\tr\left(\CG(\Delta^{(\delta)}_\varphi)-\tfrac{1}{2}\CG(\Delta^{(\delta)}_\varphi)\CG(\Delta^{(\delta)}_\varphi)\right)~,
\end{equation}
where $\Delta^{(\delta)}_\varphi$ is given by
\begin{equation}
 \Delta^{(\delta)}_\varphi:=\Delta^{(1)}_\varphi+\Delta^{(2)}_\varphi+\Delta^{(J)}_\varphi-\sum_{i,a,b,L/R}y^{i,a}_{L/R}y^{i,b}_{L/R}\tau^{L/R}_{aa}\tau^{L/R}_{bb}~.
\end{equation}
The form of the propagator $\CG$ defined in \eqref{def:propagator} now depends on the representation of the field $\varphi$. The nonvanishing components of $\CG$ for a field $\varphi$ in the various representations are: 
\begin{equation}\label{eq:G}
 \CG=\left\{\begin{array}{ll}
             \CG^{LL,LL}_{ab,cd}=\dfrac{\delta_{ac}\delta_{bd}}{p^2-|\vec{y}\,^c-\vec{y}\,^d|^2}~,&\mbox{$\varphi$ in the adjoint of $\sU(N_L)$}~,\\[0.4cm]
\CG^{RR,RR}_{ab,cd}=\dfrac{\delta_{ac}\delta_{bd}}{p^2-|\vec{y}\,^c-\vec{y}\,^d|^2}~,&\mbox{$\varphi$ in the adjoint of $\sU(N_R)$}~,\\[0.4cm]
\CG^{LR,LR}_{ab,cd}=\dfrac{\delta_{ac}\delta_{bd}}{p^2-|\vec{y}\,^c-\vec{y}\,^d|^2}~,&\mbox{$\varphi$ in the $(\overline{N}_R, N_L)$ of $\sU(N_R)\times\sU(N_L)$}~,\\[0.4cm]
             \CG_{ab}=\dfrac{\delta_{ab}}{p^2-|\vec{y}\,^a|^2}~,&\mbox{$\varphi$ in the fundamental representation}~,\\
            \end{array}\right.
\end{equation}
where the superscripts $L/R$ indicate to which gauge group the respective indices belong. Symmetry considerations now allow us to reduce \eqref{eq:1L-1} significantly. First, the term linear in $\Delta^{(J)}_\varphi$ vanishes, since $\tr(J^{MN}_\varphi)=0$. Most important, however, is  supersymmetry: Those contributions to $\delta S_{1\ell}$ which do not contain $\Delta^{(J)}_\varphi$ are, up to the factors of $\pi_\varphi,$ identical for all fields. They therefore cancel due to the relations
\begin{equation}
 6 \pi_{\CA}+4 \pi_\psi+ \pi_{gh}=0~~\eand~~\pi_{h}+2 \pi_{\chi}=0~,
\end{equation}
which reduces \eqref{eq:1L-1} to
\begin{equation}
 \delta S_{1\ell}=\sum_\varphi  \pi_\varphi \tr\left(-\tfrac{1}{2}\CG\Delta^{(J)}_\varphi\CG\Delta^{(J)}_\varphi\right)~.
\end{equation}
This expression can now be evaluated using the explicit form of the kernels $\Delta^{(J)}_\varphi$ given in \eqref{eq:Deltas1} and the propagators given in \eqref{eq:G}.

For a field $\varphi$ in the fundamental representation\footnote{The contribution for a field in the fundamental representation of $\sU(N_R)$ is the same up to replacing $L$ with $R$ everywhere.}  of $\sU(N_L)$, we arrive at the following expression in the momentum space:
\begin{equation}\label{eq:1Lfundamental}
\begin{aligned}
 \delta S_{1\ell,\varphi}=&-\tfrac{1}{2}\int \frac{\dd^3 k}{(2\pi)^3}\sum_{a,b=1}^{N_L}\mathring{\CF}^{L,a}_{MN}(k)\mathring{\CF}^{L,b}_{RS}(-k) \pi_\varphi \tr(J^{MN}_\varphi J^{RS}_\varphi)\\
&\hspace{1.0cm}\times \tr\left(\int\frac{\dd^3 p}{(2\pi)^3}\frac{\delta_{ba}}{p^2-|\vec{y}\,^{a}|^2}\tau_{aa}\frac{\delta_{ab}}{(p+k)^2-|\vec{y}\,^b|^2}\tau_{bb}\right)\\
=&-\tfrac{1}{2} \pi_\varphi \tr(J^{MN}_\varphi J^{RS}_\varphi)\int \frac{\dd^3 k}{(2\pi)^3}\sum_{a=1}^{N_L}\left(\frac{\di}{8\pi|\vec{y}\,^{a}|}+\CO(k^2)\right)\mathring{\CF}^{a,L}_{MN}(k)\mathring{\CF}^{a,L}_{RS}(-k)~,
\end{aligned}
\end{equation}
where
\begin{equation}
\begin{aligned}
&\,\tr(J^{MN}_\varphi J^{RS}_\varphi)=(g^{MR}g^{NS}-g^{MS}g^{NR})C_\varphi~,\\&C_\CA=2~,~~~C_\psi=C_\chi=1~,~~~C_{\rm gh}=C_h=0~.
\end{aligned}
\end{equation}
Fields in the adjoint representation of $\sU(N_L)$ yield the following contribution:
\begin{equation}\label{eq:1Ladjoint}
\begin{aligned}
 \delta S_{1\ell,\varphi}=&-\tfrac{1}{2}\int \frac{\dd^3 k}{(2\pi)^3}\sum_{a,b=1}^{N_L}\mathring{\CF}^{L,a}_{MN}(k)\mathring{\CF}^{L,b}_{RS}(-k) \pi_\varphi \tr(J^{MN}_\varphi J^{RS}_\varphi)\\
&\hspace{1.0cm}\times \sum_{c,d}\left(\int\frac{\dd^3 p}{(2\pi)^3}\frac{\delta_{ac}-\delta_{ad}}{p^2-|\vec{y}\,^{c}-\vec{y}\,^{d}|^2}\frac{\delta_{bc}-\delta_{bd}}{(p+k)^2-|\vec{y}\,^{c}-\vec{y}\,^{d}|^2}\right)\\
=&-\tfrac{1}{2} \pi_\varphi \tr(J^{MN}_\varphi J^{RS}_\varphi)\int \frac{\dd^3 k}{(2\pi)^3}\sum_{a,b=1,a\neq b}^{N_L}\left(\frac{\di}{8\pi|\vec{y}\,^{a}-\vec{y}\,^{b}|}+\CO(k^2)\right)\\
&\hspace{3.0cm}\times(\mathring{\CF}^{L,a}_{MN}(k)-\mathring{\CF}^{L,b}_{MN}(k))(\mathring{\CF}^{L,a}_{RS}(-k)-\mathring{\CF}^{L,b}_{RS}(-k))~.
\end{aligned}
\end{equation}
From \eqref{eq:G} it is clear that the contribution of a bifundamental field is 
\begin{multline}\label{eq:1Lbifundamental}
 \delta S_{1\ell,\varphi}=-\tfrac{1}{2}\int \frac{\dd^3 k}{(2\pi)^3}\sum_{a=1}^{N_L}\sum_{b=1}^{N_R}\mathring{\CF}^{L,a}_{MN}(k)\mathring{\CF}^{R,b}_{RS}(-k) \pi_\varphi \tr(J^{MN}_\varphi J^{RS}_\varphi)\\
\hspace{1.0cm}\times \sum_{c=1}^{N_L}\sum_{d=1}^{N_R}\left(\int\frac{\dd^3 p}{(2\pi)^3}\frac{\delta_{ac}}{p^2-|\vec{y}\,^{c}-\vec{y}\,^{d}|^2}\frac{-\delta_{bd}}{(p+k)^2-|\vec{y}\,^{c}-\vec{y}\,^{d}|^2}\right)\\
=\tfrac{1}{2} \pi_\varphi \tr(J^{MN}_\varphi J^{RS}_\varphi)\int \frac{\dd^3 k}{(2\pi)^3}\sum_{a=1}^{N_L}\sum_{b=1}^{N_R}\left(\frac{\di}{8\pi|\vec{y}\,^{a}-\vec{y}\,^{b}|}+\CO(k^2)\right)\mathring{\CF}^{L,a}_{MN}(k)\mathring{\CF}^{R,b}_{RS}(-k)~.
\end{multline}

Let us now consider the case of nonvanishing mass terms for the hypermultiplets. First, note that the cancellations we observed above are independent of the masses. (The supersymmetry cancellation is due to $\CN_{\rm 3d}=2$ SUSY, which is compatible with the introduction of mass terms.) Second, the one-loop corrections due to hypermultiplets are due to the spinors $\chi$, since the generators $J^{MN}_\varphi$ for scalars vanish. In our discussion of the superspace action in the previous sections, we introduced an $\sSO(3)$-multiplet of mass terms: a complex mass from a mass term in the superpotential, and a third real mass from adding terms to the vector superfield which is necessary for the $so(3)_Z$ R-symmetry. Their effect on our formulas is the simple shift
\begin{equation}
 |\vec{y}\,^a|\rightarrow \sqrt{(y^{1,a}-\nu^a_1)^2+(y^{2,a}-\nu^a_2)^2+(y^{3,a}-\nu^a_3)^2}=|\vec{y}\,^a-\vec{\nu}\,^a|~,
\end{equation}
where $\vec{\nu}^a=\vec{\nu}$ for all $a.$ Similarly, we could introduce mass terms for adjoint and bifundamental hypermultiplets. Their effects are slightly more complicated. As we do not need them in the subsequent discussion, we refrain from presenting them in detail.

\subsection{Special cases}

We now compare our results to those obtained in \cite{Dorey:1997ij,Dorey:1998kq}. That is, we restrict ourselves to gauge group $\sSU(2)$. A Cartan subalgebra of $\asu(2)$ can be spanned by $\sigma^3$, the third Pauli matrix, and we have
\begin{equation}
 \mathring{\CF}_{MN}=\left(\begin{array}{cc}\mathring{\CF}^1_{MN} & 0 \\ 0 & \mathring{\CF}^2_{MN}\end{array}\right)=(\mathring{\CF}^1_{MN}+\mathring{\CF}^2_{MN})\frac{\unit}{2}+(\mathring{\CF}^1_{MN}-\mathring{\CF}^2_{MN})\frac{\sigma^3}{2}=\mathring{\CF}^\unit_{MN}\frac{\unit}{2}+\mathring{\CF}^{\sigma^3}_{MN}\frac{\sigma^3}{2}~.
\end{equation}
Thus, to restrict to gauge group $\sSU(2)$, we require that $\CF^1_{MN}=-\CF^2_{MN}$. Furthermore, we can use R-symmetry to rotate the vev into the component $Y^3$. That is, $y^{1,a}=y^{2,a}=0$ and $y^{3,1}=-y^{3,2}$. 

We begin by considering the pure gauge theory without matter fields. To make contact with \cite{Dorey:1997ij}, we also introduce $M_W:=\tfrac{1}{2}y^{3,1}=-\tfrac{1}{2}y^{3,2}$. The one-loop corrections \eqref{eq:1Ladjoint} of the fields $\CA$ and $\eta$ read here as
\begin{equation}
\begin{aligned}
 \delta S_{1\ell,\CA}&=2\frac{\di}{8\pi M_W}\int \frac{\dd^3 k}{(2\pi)^3}\mathring{\CF}^{\sigma^3}_{MN}\mathring{\CF}^{\sigma^3,MN}~,\\
 \delta S_{1\ell,\eta}&=-\frac{\di}{8\pi M_W}\int \frac{\dd^3 k}{(2\pi)^3}\mathring{\CF}^{\sigma^3}_{MN}\mathring{\CF}^{\sigma^3,MN}~,
\end{aligned}
\end{equation}
which is the same result as in \cite{Dorey:1997ij}. For an adjoint hypermultiplet with mass $m_{\rm ad}$, Eq.~\eqref{eq:1Ladjoint} yields 
\begin{equation}
 \delta S_{1\ell,{\rm hyper,ad}}=-\frac{1}{2}\left(\frac{\di}{8\pi |m_{\rm ad}+M_W|}+\frac{\di}{8\pi |m_{\rm ad}-M_W|}\right)\int \frac{\dd^3 k}{(2\pi)^3}\mathring{\CF}^{\sigma^3}_{MN}\mathring{\CF}^{\sigma^3,MN}~,
\end{equation}
and for a fundamental multiplet with mass $\nu_{\rm f}$, we obtain 
\begin{equation}
 \delta S_{1\ell,{\rm hyper,f}}=-\frac{1}{8}\left(\frac{\di}{8\pi |\nu_{\rm f}+M_W|}+\frac{\di}{8\pi |\nu_{\rm f}-M_W|}\right)\int \frac{\dd^3 k}{(2\pi)^3}\mathring{\CF}^{\sigma^3}_{MN}\mathring{\CF}^{\sigma^3,MN}
\end{equation}
from Eq.~\eqref{eq:1Lfundamental}, in agreement with \cite{Dorey:1998kq}. As observed there, the $\CN_{\rm 3d}=8$ theory corresponds to one massless hypermultiplet in the adjoint representation. The hypermultiplet's contribution cancels exactly the contribution from the vector multiplet and the one-loop corrections vanish.

\subsection{The asymptotic metric on the moduli space}

The mirror gauge theory we study is given by maximally supersymmetric Yang-Mills theory in the four-dimensional bulk with $n$ bifundamental walls positioned at $s=\lambda_j$ and with $k$ fundamental walls positioned at $s=p_\sigma.$  The direction transverse to the walls is periodic.   Each fundamental multiplet confined to the wall at $s=p_\sigma$ has a mass $\vec{\nu}_\sigma.$  The gauge group on the $i^\text{th}$ interval $[\lambda_i, \lambda_{i+1}]$ is $U(N_i).$ As discussed in Section~\ref{BifundDeffects}, the space-time is cut into two halves at a point $\lambda_i$.  One gauge group $U(N_{i-1})$ acts on the boundary of the left half and another gauge group $U(N_i)$ acts on the boundary of the right half.  So far we focused on a product  $U(N_{i-1})\times U(N_{i})$ of two neighboring groups.  Here we assemble all of the contributions to extract the asymptotic metric on the Coulomb branch of this gauge theory.  To be exact, we have computed  one-loop-corrected gauge couplings which, via supersymmetry, or equivalently via hyperk\"ahlerity, completely fix the rest of the metric.  

We can read off the desired one-loop corrections from the results obtained in section \ref{Sec:Corr}. The gauge group in the $i^\text{th}$ interval between $s=\lambda_i$ and $s=\lambda_{i+1}$ is $\sU(N_i)$ and we have $n$ intervals altogether. This yields $N_1+\ldots+N_n$ Cartan generators in total, and we label them by consecutive integers. The Cartan generators of the $i^\text{th}$ interval  correspond to the integers $\mathfrak{l}_i,\ldots,\mathfrak{r}_i$, where $\mathfrak{l}_i=1+\sum_{j=1}^{i-1}N_j$ and $\mathfrak{r}_i=\sum_{j=1}^{i}N_j$. The tree-level action is
\begin{equation}
 -\frac{\di}{4}\sum_{a=1}^{N_1+\ldots+N_L}\frac{\lambda_{i+1}-\lambda_i}{g^2} \int \frac{\dd^3 k}{(2\pi)^3}
 \mathring{\CF}^a_{MN}(k)\mathring{\CF}^{a,MN}(-k)~.
\end{equation}
For the vector multiplet in the interval $i$, we obtain a correction
\begin{equation}
\begin{aligned}
\delta S_{1\ell,{\rm vector}}=\frac{\di}{16\pi}\int &\frac{\dd^3 k}{(2\pi)^3}\sum_{a,b=\mathfrak{l}_i,a\neq b}^{\mathfrak{r}_i}\frac{1}{|\vec{y}\,^a-\vec{y}\,^b|}\times\\
&\times\left(\mathring{\CF}^a_{MN}\mathring{\CF}^{a,MN}-2\mathring{\CF}^a_{MN}\mathring{\CF}^{b,MN}+\mathring{\CF}^b_{MN}\mathring{\CF}^{b,MN}\right)~.
\end{aligned}
\end{equation}
The bifundamental fields transforming nontrivially under the $U(N_i)$ factor of the gauge group are positioned at $\lambda_j$ with $j=i+1$ or $j=i$ and yield a contribution
\begin{equation}
\delta S_{1\ell,{\rm bifund.}}=-\frac{\di}{16\pi}\int \frac{\dd^3 k}{(2\pi)^3}\sum_{a=\mathfrak{l}_i}^{\mathfrak{r}_i}\sum_{b=\mathfrak{l}_j}^{\mathfrak{r}_j}\frac{1}{|\vec{y}\,^a-\vec{y}\,^b|}\left(\mathring{\CF}^a_{MN}\mathring{\CF}^{b,MN}\right)~,
\end{equation}
and each  massive hypermultiplet at $s=p_\sigma$ with $p_\sigma\in(\lambda_i,\lambda_{i+1})$ adds
\begin{equation}
\delta S_{1\ell,{\rm hyper}}=-\frac{\di}{16\pi}\int \frac{\dd^3 k}{(2\pi)^3}\sum_{a=l_1}^{r_1}\frac{1}{|\vec{y}\,^{a}-\vec{\nu}_\sigma|}\mathring{\CF}^a_{MN}\mathring{\CF}^{a,MN}~.
\end{equation}
Altogether, we get the following one-loop-corrected coupling constant for any pair of  the Cartan generators:
\begin{equation}
 \left(\frac{1}{g_{\rm 3d}^2}\right)_{ab}=\underbrace{\frac{\lambda_{i+1}-\lambda_i}{g^2}\delta_{ab}}_{\mbox{tree level}}+\underbrace{s_{ab}}_{\mbox{vector mplt.}}+\underbrace{n_{ab}}_{\mbox{bifundamentals}}+\underbrace{d_{ab}}_{\mbox{fundamentals}}~,
\end{equation}
where
\begin{equation}
\begin{aligned}
 s_{ab}&=\left\{\begin{array}{ll}
-\sum_{c=\mathfrak{l}_i}^{\mathfrak{r}_i}\frac{1}{2\pi|\vec{y}\,^{a}-\vec{y}\,^{c}|}~&~\mbox{for}~a=b~,~a\in[\mathfrak{l}_i,\mathfrak{r}_i]~,\\
\frac{1}{2\pi|\vec{y}\,^{a}-\vec{y}\,^{b}|}~&~\mbox{for}~a\neq b~,~a,b\in[\mathfrak{l}_i,\mathfrak{r}_i]~,\\
0~&~\mbox{otherwise}~,
\end{array}\right.\\
 n_{ab}&=\left\{\begin{array}{ll}
\frac{1}{4\pi|\vec{y}\,^{a}-\vec{y}\,^{b}|}~&~\mbox{for}~a\in[\mathfrak{l}_i,\mathfrak{r}_i]~,~b\in[\mathfrak{l}_j,\mathfrak{r}_j]~,~i=j\pm 1~,\\
0~&~\mbox{otherwise}~,
\end{array}\right.\\
 d_{ab}&=\left\{\begin{array}{ll}
\sum_{\sigma|\lambda_i<p_\sigma<\lambda_{i+1}}\frac{1}{4\pi|\vec{y}\,^{a}-\vec{m}|}~&~\mbox{for}~a=b~,~a\in[\mathfrak{l}_1,\mathfrak{r}_1]~,\\
0~&~\mbox{otherwise}~.
\end{array}\right.
\end{aligned}
\end{equation}
Our result agrees with the asymptotic metric for balanced representations of $A_k$ bows as given in \cite{Cherkis:2010bn}, Eqns.\ (128)-(130).

\section{Conclusions}\label{sec:Conclusion}

While gauge theories with impurities are interesting in their own right, here we employed them as a tool for studying the moduli spaces of instantons on multi-Taub-NUT spaces. To this end we used maximally supersymmetric four-dimensional Yang-Mills theory coupled to both fundamental and bifundamental $\CN_{\rm 3d}=4$ matter confined to three-dimensional impurity walls. Just as in \cite{DeWolfe:2001pq,Erdmenger:2002ex}, we worked in $\CN_{\rm 3d}=2$ superspace language. This allowed us to study the moduli space of vacua of the gauge theory by considering D- and F-flatness conditions.

A string theory realization of this theory is given by the Chalmers-Hanany-Witten configuration of branes in type IIB string theory. Via T-duality, this configuration is related to another string theory background which can be effectively described by Yang-Mills instantons on multi-Taub-NUT space $TN_k$. Here, the number $k$ of the Taub-NUT centers corresponds to the number of the NS5 branes in the CHW configuration.  It is the Higgs branch of the impurity gauge theory that is identified via T-duality as the moduli space of instantons on $TN_k$.

To identify the Higgs branch, we derived conditions for vacuum configurations. We found that the resulting equations are exactly the moment map conditions appearing in the bow construction of \cite{Cherkis:2009jm} and \cite{Cherkis:2010bn}.  This independently verifies the string duality statement.  
	
We then used this relation to compute the asymptotic metric of the moduli space of instantons on $TN_k$ using the gauge theory as follows:  Applying electric-magnetic duality to our impurity theory, we obtain the same type of gauge theory with  the two types of impurity walls interchanged.  The resulting mirror theory has the moduli space of instantons on $TN_k$ as its Coulomb branch.  Its metric is determined by the kinetic term couplings of the effective theory.  We performed a one-loop background field computation, which is an extension of the calculation presented in \cite{Dorey:1997ij,Dorey:1998kq}. The resulting asymptotic metric on the Coulomb branch is exactly the asymptotic metric on the moduli space of instantons found in \cite{Cherkis:2010bn}. 

We expect that the techniques we have used here can be fruitfully applied to other questions. For example, the superfield action we have used defines a gauge theory with impurities for any bow representation.  One could use this theory to verify that the moduli space of an E-type bow is insensitive to the interval lengths (i.e.\ to the three-dimensional couplings) and that it coincides with the moduli space of a quiver representation obtained by shrinking all of the bow intervals to zero size.  Another intriguing direction for future research is to explore such impurity theories on curved space-time background.

\section*{Acknowledgements}

We thank Martin Ro\v{c}ek for his comments about the manuscript.  SCh is grateful to Amihay Hanany for useful discussions. The work of CS was supported by a Career Acceleration Fellowship from the UK Engineering and Physical Sciences Research Council.

\appendix

\section*{Appendix}

We use the embedding of $\CN_{\rm 3d}=2$ superspace into $\CN_{d=4}=2$ superspace as presented in \cite{Erdmenger:2002ex}. The coordinates $(x^{\hat{\mu}},\theta^1,\bar\theta_1,\theta^2,\bar\theta_2)$, where $\hat{\mu}= \hat{0},\hat{1},\hat{3},$ parameterize the $\CN_{\rm 4d}=2$ superspace.  Here $x^0=x^{\hat{0}}, x^1=x^{\hat{1}}, $ and $x^2=x^{\hat{3}}.$  Consider the linear combinations
\begin{equation}\label{eq:twistGrassmann}
 \theta=\tfrac{1}{2}(\theta_1+\bth^1-\theta_2-\bth^2)\ \eand\  \thetas=\tfrac{1}{2\di}(\theta_1-\bth^1-\theta_2+\bth^2)~,
\end{equation}
with analogous linear combinations for the supercharges and the superspace covariant derivatives. The subspace given by $\thetas=0$ and  $x^{\hat{2}}=s,$ with some fixed value of the parameter $s,$ is then preserved by the $\CN_{\rm 3d}=2$ supersymmetry algebra. Thus one can use the coordinates $(x^\mu,\theta,\bar{\theta}), \mu=0,1,2,$ to parameterize the three-dimensional superspace, with $s$ playing the role of a parameter external to this superspace.  

We take our fields to be antihermitian so that $F_{\mu\nu}:=\partial_\mu A_\nu-\partial_\nu A_\mu+[A_\mu,A_\nu]$ and the covariant derivatives are of the form $\nabla_\mu B=\partial_\mu B+[A_\mu,B]$ or $\nabla_\mu Q=\partial_\mu Q+A_\mu Q$. A bar denotes hermitian conjugation.

\subsection*{Spinor conventions}

We use the standard superfield conventions as given, e.g., in \cite{Wess:1992cp}. The metric is $\eta_{\mu\nu}=$diag$(-1,1,1,1)$. We use $\epsilon^{12}=-\epsilon_{12}=1$ for raising and lowering spinor indices. So $\psi^\alpha=\epsilon^{\alpha\beta}\psi_\beta$ and $\psi_\alpha=\epsilon_{\alpha\beta}\psi^\beta$, where $\epsilon^{\alpha\beta}=\di\sigma^2$. The Pauli matrices are
$\sigma^1=\left(\begin{smallmatrix}
0&1\\
1&0\end{smallmatrix}\right), \ 
\sigma^2=\left(\begin{smallmatrix} 0&-\di\\\di&0\end{smallmatrix}\right), \ 
\sigma^3=\left(\begin{smallmatrix}
1&0\\
0&-1\end{smallmatrix}\right)$
and $\sigma_0=-1_{2\times2}.$ 
The spinor summation conventions are
\begin{equation}
\psi\chi=\psi^\alpha\chi_\alpha=-\psi_\alpha\chi^\alpha=\chi^\alpha\psi_\alpha=\chi\psi~,
\end{equation}
\begin{equation}
\bar{\psi}\bar{\chi}=\bar{\psi}_\alpha\bar{\chi}^\alpha=-\bar{\psi}^\alpha\bar{\chi}_\alpha=\bar{\chi}_\alpha\bar{\psi}^\alpha=\bar{\chi}\bar{\psi}~,
\end{equation}
and we have
\begin{equation}\label{spincon3}
\psi^\alpha\bar{\chi}_\alpha=-\psi_\alpha\bar{\chi}^\alpha=\bar{\chi}^\alpha\psi_\alpha~.
\end{equation}
Some useful spinor relations are:
\begin{align}\label{spinrel}
\theta^\alpha\theta^\beta&=-\tfrac{1}{2}\epsilon^{\alpha\beta}\theta^2~,&
\bar{\theta}^\alpha\bar{\theta}^\beta&=\tfrac{1}{2}\epsilon^{\alpha\beta}\bar{\theta}^2~,\\
\theta_\alpha\theta^\beta&=-\tfrac{1}{2}\delta_\alpha^\beta\theta^2~,&
(\theta\sigma^{\hat{\mu}}\bar{\theta})(\theta\sigma^{\hat{\mu}}\bar{\theta})&=-\tfrac{1}{2}\theta^2\bar{\theta}^2\eta^{{\hat{\mu}}{\hat{\nu}}}~.
\end{align}
In particular for Equation (4) in Section 3.1, we have 
\begin{equation}
\begin{split}
(\bar{\theta}_\alpha\theta^\alpha)^2=& \bar{\theta}_\alpha\theta^\alpha\bar{\theta}_\beta\theta^\beta =-\theta^\alpha\bar{\theta}_\alpha\bar{\theta}_\beta\theta^\beta \\
=& \tfrac{1}{2}\bar{\theta}^2\theta^\alpha\epsilon_{\alpha\beta}\theta^\beta = \tfrac{1}{2}\bar{\theta}^2\theta^\alpha\theta_\alpha = \tfrac{1}{2}\bar{\theta}^2\theta^2~,
\end{split}
\end{equation}
where we have used Equation $(\ref{spincon3})$ in the first line and Equation ($\ref{spinrel}$) in going from the first to the second line.

Integration in superspace has the following properties:
$
\int d\theta=0~,\
\int d\theta\  \theta=1~,$
so that
$
\int d\theta_\alpha \theta^\beta=\partial_\alpha\theta^\beta=\delta_\alpha^\beta~.
$

\subsection*{Superfields}

The $\CN_{\rm 3d}=2$ chiral superfields read in chiral coordinates $y^{\hat{\mu}}=x^{\hat{\mu}}+\di\theta\sigma^{\hat{\mu}}\bth$, $\hat{\mu}=\hat{0},\hat{1},\hat{3}$, as follows:
\begin{equation}
\begin{aligned}
 \CX&=v_6(y)+\di Y_1(y)+\sqrt{2}\theta\psi(y)+\theta^2 G(y)~,\\
 \CU^1&=\CZ=Z(y)+\sqrt{2}\theta\chi^i(y)+\theta^2 F^i(y)~,\\
 \CU^2&=\CY=Y(y)+\sqrt{2}\theta\chi^i(y)+\theta^2 F^i(y)~.
\end{aligned}
\end{equation}
while the $\CN_{\rm 3d}=2$ vector superfield is given in Wess-Zumino gauge by 
\begin{equation}
 \CV=-\theta\sigma^2\bth Z_3-\theta\sigma^\mu\bth v_\mu+\di\theta^2\bth\bar{\lambda}-\di\bth^2\theta\lambda+\tfrac{1}{2}\theta^2\bth^2D~.
\end{equation}
The vector superfield $\CV$ together with the chiral superfield $\CX$ form a $\CN_{\rm 4d}=2$ vector supermultiplet:
\begin{equation}
 \Psi(y)=\CX(y,\theta_1)+\di \sqrt{2}\theta_2W(y,\theta_1)+\theta_2^2G(Y,\theta_1)~.
\end{equation}
Here, $W_\alpha$ is the chiral fermionic field strength of the superfield $\CV$. Performing now the coordinate transformation \eqref{eq:twistGrassmann}, the $\CN_{\rm 4d}=2$ vector superfield contains at $\thetas=0$ an $\CN_{\rm 3d}=2$ chiral superfield $\CX$ and the $\CN_{\rm 3d}=2$ complex linear superfield $\Sigma$.

\subsubsection*{The linear multiplet}

The vector multiplet gives rise to the linear multiplet $\Sigma$ defined by 
\begin{equation}
\Sigma=\epsilon^{\alpha\beta}\bar{D}_\alpha(e^{2\di \CV}D_\beta e^{-2\di \CV})~.
\end{equation}
The calculation is made simpler by writing $\Sigma$ as a function of $y^{\hat{\mu}}=x^{\hat{\mu}}+\di\theta\sigma^{\hat{\mu}}\bar{\theta}$:
\begin{equation}
\CV(y) = -\di \theta \bar{\theta} Z_3 - \theta \sigma^{\hat{\mu}} \bar{\theta} v_{\hat{\mu}} + \di \theta^2 \bar{\theta} \bar{\lambda} - \di \bar{\theta}^2  \theta \lambda + \frac{1}{2} \theta^2 \bar{\theta}^2 (D-\di\partial_{\hat{\mu}} v^{\hat{\mu}})~. 
\end{equation}
We note that 
\begin{equation}\label{linearmult}
e^{2\di\CV}D_\beta e^{-2\di\CV}=-2\di D_\beta \CV+2[\CV, D_\beta \CV]~,
\end{equation} 
since powers of $\CV$ higher than $\CV^2$ vanish due to the properties of the Gra{\ss}mann variables $\theta$ and $\bar{\theta}$. Writing $D_\beta \CV$ in components:
\begin{equation}
\begin{split}
D_\beta \CV=&(\partial_\beta+2\di\sigma^{\hat{\mu}}_{\beta\gamma}\bar{\theta}^\gamma\partial_{\hat{\mu}})\CV(y)\\
=&-\di\bar{\theta}_\beta Z_3-\sigma^{\hat{\mu}}_{\beta\gamma}\bar{\theta}^\gamma v_{\hat{\mu}}+2\di\theta_\beta\bar{\theta}\bar{\lambda}-\di\bar{\theta}^2 \lambda_\beta+\theta_\beta\bar{\theta}^2 D\\
&+\sigma^{\hat{\mu}}_{\beta\gamma}\theta^\gamma\bar{\theta}^2\partial_{\hat{\mu}}Z_3-\di\bar{\theta}^2\sigma^{{\hat{\mu}}{\hat{\nu}}\gamma}_\beta\theta_\gamma(\partial_{\hat{\mu}} v_{\hat{\nu}}-\partial_{\hat{\nu}} v_{\hat{\mu}})+\theta^2\bar{\theta}^2\sigma^{\hat{\mu}}_{\beta\gamma}\partial_{\hat{\mu}}\bar{\lambda}^\gamma~,
\end{split}
\end{equation}
we can now write an expression for the commutator:
\begin{align}
[\CV, D_\beta \CV]=&-\bar{\theta}^2\sigma_\beta^{{\hat{\mu}}{\hat{\nu}}\gamma}\theta_\gamma[v_{\hat{\mu}}, v_{\hat{\nu}}]-\di\theta^2\bar{\theta}^2\sigma^{\hat{\mu}}_{\beta\gamma}[v_{\hat{\mu}}, \bar{\lambda}^\gamma]~.
\end{align}
Now we have Eq.~(\ref{linearmult}) in component form and since $\Sigma$ is a function of $y$, the covariant derivative $\bar{D}_\alpha$ reduces to $-\bar{\partial}_\alpha$. Using the Taylor expansion $f(x)=f(y)-\di\theta\sigma^{\hat{\mu}}\bar{\theta}\partial_{\hat{\mu}} f(y)+\frac{1}{4}(\theta\theta)(\bar{\theta}\bar{\theta})\Box f(y)$ we retrieve the component expansion for the  linear multiplet $\Sigma(x)$:
\begin{equation}
\begin{split}
\Sigma(x)=&4Z_3-4\theta^\alpha\bar{\lambda}_\alpha-4\bar{\theta}_\alpha\lambda^\alpha-4\di\theta^\alpha\bar{\theta}_\alpha D-2\theta\sigma^{\hat{\mu}}\bar{\theta}\epsilon_{\hat{\mu}\hat{\nu}\hat{\kappa}} F^{\hat{\nu}\hat{\kappa}}-2\di\bar{\theta}^2\theta\sigma^{\hat{\mu}}\nabla_{\hat{\mu}}\lambda\\
&+2\di\theta^2\bar{\theta}\bar{\sigma}^{\hat{\mu}}\nabla_{\hat{\mu}}\bar{\lambda}-\theta^2\bar{\theta}^2\Box Z_3~,
\end{split}
\end{equation}
where $\epsilon_{\hat{0}\hat{1}\hat{3}}=-1.$

\subsubsection*{$\CN_{\rm 3d}=2$  matter supermultiplets on impurity walls}
Each fundamental defect wall is carrying a fundamental chiral supermultiplet
\begin{equation}
\CQ_{1j}=Q_{1j}+\sqrt{2}\theta\zeta_{1j}+\theta^2 J_{1j}~,
\end{equation}
and an anti-fundamental chiral supermultiplet
\begin{equation}
\CQ_{2j}=Q_{2j}+\sqrt{2}\theta\zeta_{2j}+\theta^2 J_{2j}~.
\end{equation}
Each bifundamental defect wall carries one chiral supermultiplet in the  $(N_{\sigma-1}, \bar{N}_\sigma)$ representation of 
$U(N_{\sigma-1})\times U(N_\sigma)$
\begin{equation}
\CB_{1j}=B_{1j}+\sqrt{2}\theta\xi_{1j}+\theta^2 L_{1j}~,
\end{equation}
and one chiral supermultiplet in the  $(\bar{N}_{\sigma-1}, N_\sigma)$ representation
\begin{equation}
\CB_{2j}=B_{2j}+\sqrt{2}\theta\xi_{2j}+\theta^2 L_{2j}~.
\end{equation}



\end{document}